\documentclass[journal ]{new-aiaa}
\usepackage[utf8]{inputenc}
\usepackage{textcomp}

\usepackage{graphicx}
\usepackage{amsmath}
\usepackage[version=4]{mhchem}
\usepackage{siunitx}
\usepackage{longtable,tabularx}
\usepackage{bm}
\usepackage{footmisc}
\usepackage{hyperref}
\usepackage{cases}
\usepackage{theorem}
\usepackage{cleveref}
\usepackage{float}

\usepackage{xparse}
\ExplSyntaxOn
\NewDocumentCommand{\mref}{m}{\quinn_mref:n {#1}}
\seq_new:N \l_quinn_mref_seq
\cs_new:Npn \quinn_mref:n #1
 {
  \seq_set_split:Nnn \l_quinn_mref_seq { , } { #1 }
  \seq_pop_right:NN \l_quinn_mref_seq \l_tmpa_tl
  ( 
  \seq_map_inline:Nn \l_quinn_mref_seq
    { \ref{##1}-\nobreakspace } 
  \exp_args:NV \ref \l_tmpa_tl 
  ) 
 }
\ExplSyntaxOff

\usepackage{xcolor}

\newtheorem{theorem}{Theorem}

\title{Full-Envelope Flight Control for Compound Vertical Takeoff and Landing Aircraft}

\author{Jean-Marie KAI\footnote{Flight Control Engineer, Energy and Propulsion Department, Safran Tech, Magny-Les-Hameaux, France, jean-marie.kai@ariane.group .}}
\affil{Safran Tech, Energy \& Propulsion Department, Magny-Les-Hameaux, France}

\begin{document}

\maketitle

\begin{abstract}

This paper presents a flight control design for compound Vertical Takeoff and Landing (VTOL) vehicles. With their multitude of degrees of controllability as well as the significant variations in their flight characteristics, VTOL vehicles present challenges when it comes to designing their flight control system, especially for the transition phase where the vehicle transitions between near-hovering and high-speed wing-borne flights. This work extends previous research on the design of unified and generic control laws that can be applied to a broad class of vehicles such as hovering vehicles and fixed-wing aircraft. This paper exploits this unifying property and presents an extension for the case of compound VTOL vehicles. The proposed control approach consists of nonlinear geometric control laws that are continuously applicable  over the entire flight envelope, excluding the use of switching policies between different control algorithms. A transition strategy consisting of a sequence of high-level setpoints is associated with the flight control laws, it is defined with respect to flight envelope limitations and is applied in this work to a commercially available compound unmanned aerial vehicle. The control algorithms are implemented on a Pixhawk controller, they are evaluated via Hardware-In-the-Loop simulations and finally validated in a flight experiment.

\end{abstract}

\section*{Nomenclature}

{\renewcommand\arraystretch{1.0}
\noindent\begin{longtable*}{@{}l @{\quad=\quad} l@{}}

${\mathcal I}=\{O, \bm \imath_0, \bm \jmath_0, \bm k_0   \}$ & inertial frame associated with a 3D Euclidean vector space ${\bm E}^3$, with $\bm k_0$ pointing downward\\
${\mathcal B}=\{G, \bm \imath, \bm \jmath, \bm k  \}$ &  body-fixed frame, with $\bm \imath$ pointing along the fuselage axis and $\bm \jmath$ pointing toward the right wing\\
$\Pi_{\bm u}$ & operator of projection on the plane orthogonal to $\bm u$\\
$x \in \mathbb{R}^3$ & coordinate vector of $\bm x \in {\bm E}^3$ in the body-fixed frame ${\mathcal B}$ \\
$x_i \ \  (i=1,2,3)$ & $i$th components of $x \in \mathbb{R}^3$, in other words, $x_1= \bm x \cdot \bm \imath$, $x_2= \bm x \cdot \bm \jmath$, $x_3= \bm x \cdot \bm k$ \\
$S(.)$ & Skew symmetric operator associated with the cross product of vectors, $S(x)y= x \times y$ \\
$\bm v$ & velocity of the center of mass with respect to ${\mathcal I}$, $m/s$\\
$\bm v_w$ & ambient wind velocity with respect to ${\mathcal I}$, $m/s$\\
$\bm v_a =\bm v - \bm v_w$ & aircraft air velocity with respect to ${\mathcal I}$, $m/s$ \\
$m$ & vehicle's mass, $Kg$ \\
$g_0$ & gravity acceleration, $m/s^2$\\
$\rho$ & air density, $Kg/m^3$\\
$S$ & reference area, $m^2$\\
$b$ & lateral reference length, $m$\\
$c$ & longitudinal reference length, $m$\\
$C_D$ & drag coefficient\\
$C_L$ & lift coefficient\\
$C_{D,0}$ & zero-lift drag coefficient\\
$C_{L,\alpha}$ & lift coefficient slope, $rad^{-1}$\\
$C_{Y,\beta}$ & side force coefficient slope, $rad^{-1}$\\
\multicolumn{2}{@{}l}{Acronyms}\\
$DRB$ & Disturbance Rejection Bandwidth\\
$DRP$ & Disturbance Rejection Peak\\
\multicolumn{2}{@{}l}{Subscripts}\\
$FW$ & fixed-wing configuration related variable\\
$MC$ & multicopter configuration related variable\\

\end{longtable*}}

\section{Introduction}

\lettrine{N}{ovel} and unconventional Vertical Take-off and Landing (VTOL) configurations are emerging, thanks to advances in electric propulsion technologies, battery technologies, and embedded computer systems. Their development is accelerated by the growing needs related to Unmanned Aerial Vehicles (UAV) and Urban Air Mobility (UAM) applications. These VTOL configurations offer the advantage of not requiring any runway but only a helipad, or even a simple flat terrain in the case of military applications. 
The trend observed today is to design VTOL vehicles, which after vertical take-off, are able to reconfigure towards high-speed cruise flights that rely on aerodynamic lift. This is driven by the superior efficiency demonstrated by fixed-wing aircraft compared to their rotary-wing counterparts. This has led to the design of wing-equipped VTOL vehicles in a diverse range of configurations, among which we can cite the compound configuration (sometimes referred to as lift plus cruise) which uses two different fixed propulsion systems for hover and cruise flights, the tilt-rotor configurations where the propulsion systems are mounted on rotating shafts and contribute to both hover and cruise flights, and finally the tilt-wing configuration where the main wing tilts along with the propelling effectors  \cite{configs,trendsVTOL}.

Because of their wide flight envelope, these VTOL vehicles present design challenges, especially when it comes to designing flight control laws. Indeed, the control system has to deal with near hovering flights, cruise flights  and transition maneuvers between them. Historically, methods for the development of control systems for an airplane or for a rotorcraft (helicopters, for instance) have diverged in different ways, by associating control variables differently with the state variables to be controlled and by employing different control theories. Today, the techniques for controlling a fixed-wing aircraft and a rotary-wing aircraft are essentially different. For UAV applications, the first methods that were explored to command VTOL configurations consisted in running in parallel two different control applications for the quasi-stationary and cruising modes and blending their outputs adequately during the transition phases. This simple approach is widely used and can be found in open-source UAV autopilot software \cite{analysis-px4}. However, although this technique can be employed for small vehicles that have a high thrust-to-weight ratio in favor of enhanced robustness, its applicability to large manned vehicles is to be questioned with regards to safety and narrow flight corridor requirements and the necessity to achieve \textit{smooth} transitions for passengers comfort. 
Other methods that were applied to this control problem consist in working a set of linearized models along the full flight envelope and designing gain scheduling linear controllers; see for example Refs.~\cite{scheduling1,scheduling2,7152383}. However, these techniques have their limitations when dealing with unsteady and highly nonlinear transition maneuvers of VTOL vehicles. These challenges have driven research communities and industrialists to adopt unified control approaches suitable for VTOL configurations that do not require (or at least greatly reduce) switching between largely different control  structures.

Research on unified control concepts for VTOL vehicles dates back to the development of Vector thrust Aircraft Advanced Control (VAAC) for the AV-8B Harrier in the 1970s and 1980s, where the piloting interface was simplified thanks to novel control  strategies that would operate in a single mode throughout the entire flight phases while integrating flight and propulsion controls. This concept was then further developed and implemented in the F-35B fighter aircraft as mentioned in Ref.~\cite{lombaerts-concepts}, and detailed in Refs.~\cite{F35B1,F35B2,F35B3}. Advancements on unified control approaches are being considered today for electric VTOL (eVTOL) vehicles especially for UAM applications. In Ref.~\cite{Lombaerts-unified} for instance, flight control laws based on Incremental Nonlinear Dynamic Inversion (INDI) for internal loops, and linear controllers for external loops were designed and tested in simulation for a conceptual eVTOL configuration that has seperate lift and cruise rotors. Similar methods are applied in Refs.~\cite{tiltwing-unified,NDIatt,panish2} for a tilt-wing UAV configuration, and in Ref.~\cite{DIFRANCESCO2015156}  for a tilt-rotor UAV configuration. In Ref.~\cite{Hartmann}, a unified velocity control is applied to a tilt-wing aircraft; the methodology employs map-based feedforward controllers that maintain trimmed straight-lined flight as well as a set of virtual control surfaces that allow to transition between different flight states. The latter work is complemented in Ref.~\cite{hartmannpos} with high-level position control loops with a focus on approach and departure maneuvers including flight state transitions. Other nonlinear techniques that are employed for the control of VTOL vehicles involve neural-network based dynamic inversion such as in Refs.~\cite{NN1,NN2}, and Nonlinear Model Predictive Control (NMPC) methods such as in Ref.~\cite{ALLENSPACH2021109790}.

 The work in this paper comes as a complement to previous work, which first addressed the control of standard quadcopter configurations using geometric nonlinear methods \cite{hua2009control,kai2017quad}. With the aim of establishing a unified approach to the control of aerial vehicles, this geometric nonlinear control was then extended to axisymmetric vehicles by including a nonlinear analytical expression of aerodynamic forces in the control model \cite{pucci2015nonlinear}. The approach was extended further to fixed-wing vehicles and was validated in flight tests \cite{2018flight,2019-unified-automatica,2018thesis}. The proposed control laws in this paper inherit from the stability and convergence properties of the work in Ref.~\cite{2019-unified-automatica}, and then in Ref.~\cite{2019convertible} in which an extension to tilt-rotor convertible vehicles is proposed. This paper contributes further by extending these concepts to compound configurations. Compared to the work in Ref.~\cite{2019convertible}, modifications and improvements were made regarding, among other things, the generalization of the control loops over the entire flight envelope covering simultaneously near-hover and high-speed flights. This has eliminated the necessity to blend intermediate attitude control outputs, therefore fully exploiting the potential of the nonlinear control design. In addition, it was sufficient to design the transition and back-transition maneuvers as a sequence of high-level setpoints, which further highlights the advantages of unified control loops that become transparent to the pilot or the operator.

The work in this paper presents contributions to the development of flight control laws for VTOL aircraft and more specifically for compound VTOL configurations.  The proposed control design falls into the category of unified and nonlinear control approaches. More specifically, it is based on a common and generic nonlinear control design that can be applied to a broad class of vehicles such as rotorcrafts and fixed-wing vehicles. Indeed, this is particularly useful for hybrid VTOL configurations because a common control law for the entire flight envelope has the potential to exclude switching policies between different control algorithms. These concepts are applied in this paper and are detailed for the compound VTOL configuration. The control laws described in this paper are essentially nonlinear and provide an alternative to current state-of-the-art flight control solutions such as designing linear control feedback around equilibrium trim trajectories. The nonlinear control terms are illustrated by the use of geometric feedback terms complemented by feedforward terms as well as nonlinear transformations that map desired accelerations to attitude and thrust setpoints as well as angular accelerations to torque actuation control terms. These transformations are based on an analytical physical control model that is sufficiently representative while being simple to lend itself to control design. An additional originality of these control laws is that they do not rely on the classic Euler angles (roll, pitch and yaw) nor on an angular parametrization of the air velocity such as the angle-of-attack or the side-slip angle; this allows us to overcome singularities associated with their definition at low-speed flights and offers an opportunity to apply a unique control design over a wide speed range.
Compared to other nonlinear methods such as INDI techniques, the control laws presented in this paper can be seen as a compromise between not relying on acceleration and angular acceleration estimation (therefore being less prone to noise propagation) and requiring a small set of aerodynamic coefficients. Additionally, they are capable of achieving autonomous flights in any phase or configuration by accounting for aerodynamic forces in external control loops instead of requiring a pilot to command attitude angles or angular rates.

The paper is organized as follows. Section \ref{sec:controlmodel}, presents the control model along with the different assumptions. The control structure and detailed control laws are defined in section \ref{sec:control}. The high-level transition and back-transition strategies are then presented in Sec.~\ref{sec:trans-strat}. Hardware-In-the-Loop simulation results are shown in Sec.~\ref{sec:HIL}, and flight test results are shown in Sec.~\ref{sec:test}. Finally, concluding remarks are given in Sec.~\ref{sec:conclusion}.

\section{Control Model}\label{sec:controlmodel}
\begin{figure}[hbt!]
\centering
\begin{minipage}{.45\textwidth}
\centering
\includegraphics[width=.9\textwidth]{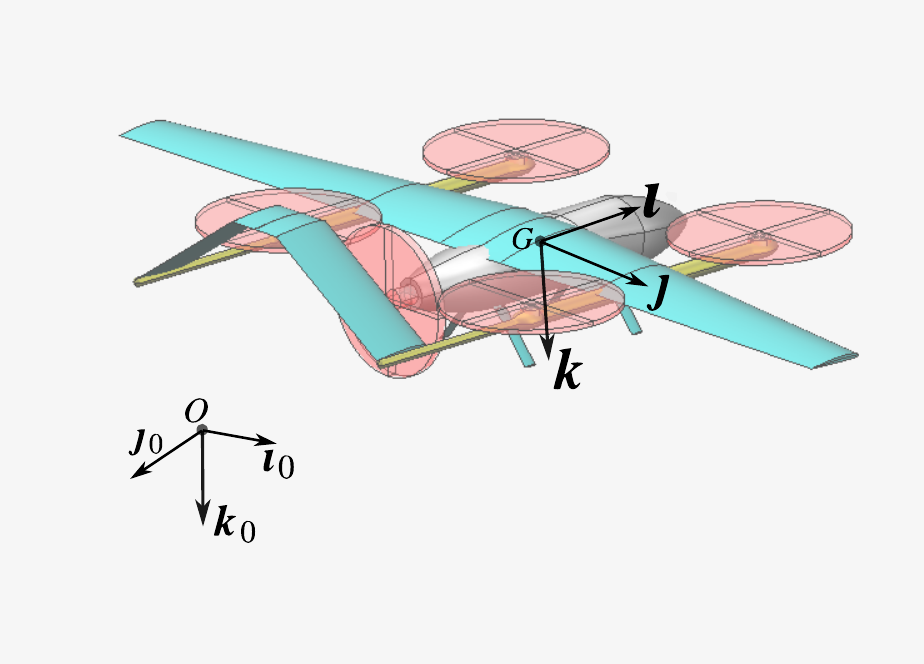}
\caption{Inertial frame and body-fixed frame axis.}
\label{fig:axis}
\end{minipage}
\begin{minipage}{.45\textwidth}
\centering
\includegraphics[width=1.0\textwidth]{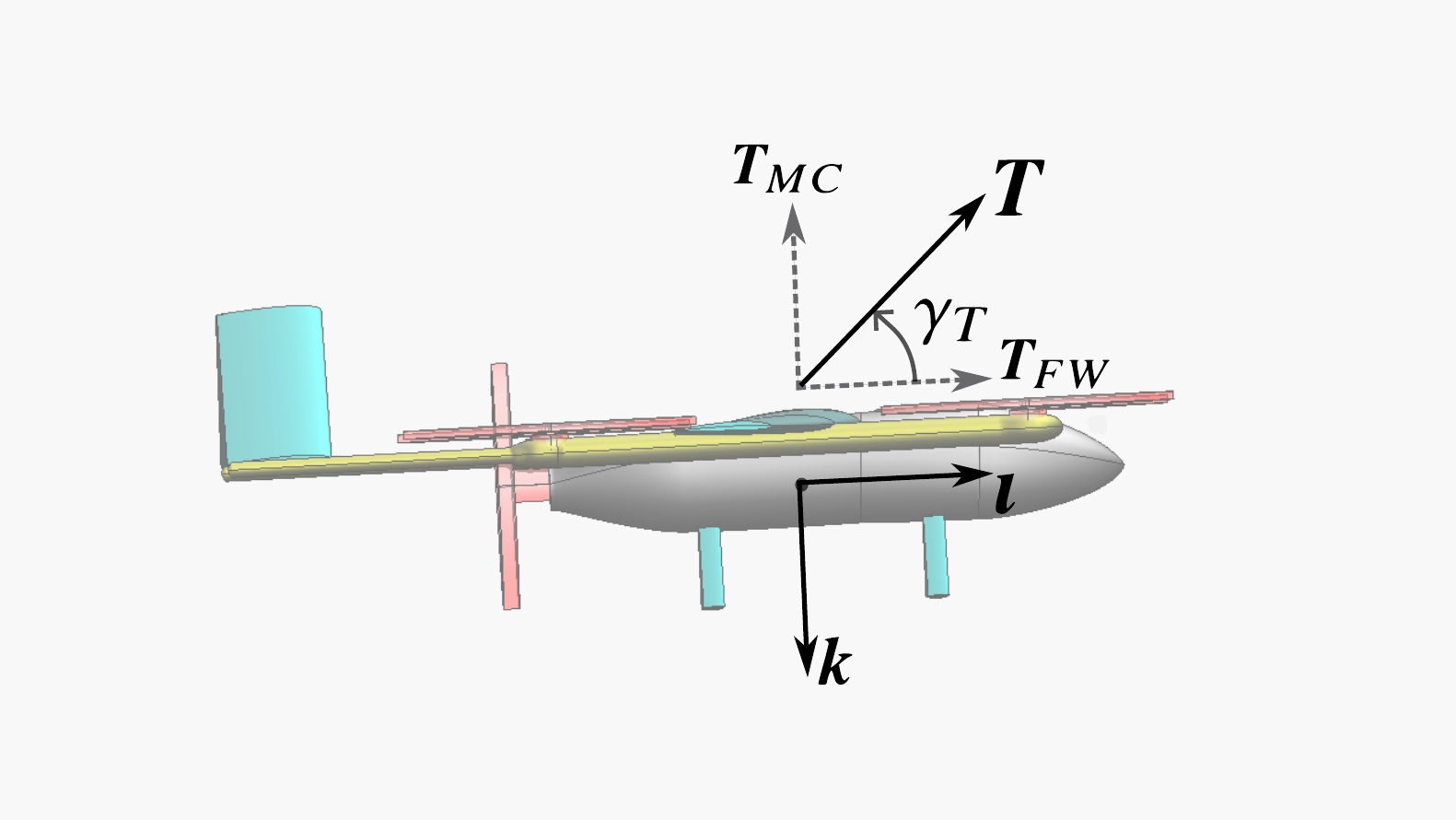}
\caption{Thrust vectoring representation.}
\label{fig:thrust-dir}
\end{minipage}
\end{figure}

The objective of this section is to derive the differential equations which govern the evolution of the system. These equations will constitute our control model, which, contrary to a detailed simulation model, is supposed to be simple without losing its relevance and will constitute a basis for the design of the control laws.
Our system is an aerial vehicle which is considered to be a rigid body; therefore, the state variables describing its motion can be reduced to the position $\bm r$ and velocity $\bm v$ of its center of mass $G$ as well as its attitude represented by the body axis $(\bm \imath,\bm \jmath,\bm k)$ (see Fig.~\ref{fig:axis}) and angular velocity $\bm \omega$.

The ordinary differential equations that describe the time evolution of these variables are obtained by applying the Newton-Euler Formalism, with the Earth-fixed frame approximated as being an inertial frame. This yields the following set of kinematic and dynamic equations:

\begin{numcases}{}
\dot{\bm r} ={\bm v}  \label{eq:trans-kin} \\
\dot{\bm v} ={\bm g} + \frac{{\bm F}_a}{m} + \frac{\bm T}{m} \label{eq:trans-dyn} \\
\frac{d}{dt}(\bm \imath, \bm \jmath, \bm k) ={\bm \omega} \times (\bm \imath, \bm \jmath, \bm k) \label{eq:rot-kin} \\
J \dot{\omega} =-S(\omega)J\omega +M_a + M  \label{eq:rot-dyn} 
\end{numcases}

Where ${\bm g}=g_0 \bm k_0$ is the gravity vector, ${\bm F}_a$ is the resultant aeordynamic force, and ${\bm T}$ is the total thrust vector generated by the rotors. $J$ is the body inertia matrix, $M$ is the control torque produced by control surfaces or by propulsion systems, and $M_a$ is the residual aerodynamic torque related to the aircraft's fixed surfaces. Equations \mref{eq:trans-kin,eq:rot-dyn} constitute the state equations for our system, with the vectors $\bm T$ and $M$ being the control terms.

\subsection{Thrust force}

 Unlike multicopter or fixed-wing configurations, hybrid VTOL aircraft enable thrust vectoring capabilities via tilt-rotor mechanisms or by being equipped with multiple rotors generating thrust forces in different directions, as is the case for compound configurations. The direction of the total thrust is an additional degree of freedom represented here by the angle $\gamma_T$ according to Fig.~\ref{fig:thrust-dir}. The total thrust force $\bm T$ can then be expressed as follows:
 \begin{equation} \label{eq:thrust-dir}
\bm T= |\bm T| (\cos(\gamma_T) \bm \imath + \sin(\gamma_T) \bm k)
\end{equation}
This mathematical representation is valid for both tilt-rotor and compound configurations. In the latter case (see Fig.~\ref{fig:thrust-dir}) the total thrust force is the sum of the pusher thrust $\bm T_{FW}$ and the collective thrust generated by the lift rotors $\bm T_{MC}$, in other words, $\bm T = \bm T_{FW} + \bm T_{MC}$. Assuming the pusher rotor is aligned with $ \bm \imath$,  the seperated thrust components can be deduced from $|\bm T|$ and $\gamma_T$ as follows:

\begin{align} \label{eq:seperated-thrust}
\begin{split}
\bm T_{MC} &=  |\bm T| \sin(\gamma_T) \bm k \\
\bm T_{FW} &=  |\bm T| \cos(\gamma_T) \bm \imath
\end{split}
\end{align}

Also note that these modeling expressions cover the special cases of multicopter vehicles by fixing $\gamma_T$ to $-\pi/2$ and fixed-wing aircraft by fixing $\gamma_T$ to 0 or to any constant value representing the actual direction of the pusher rotor. For the sake of simplicity, the normal forces that appear in the plane of propellers operating at a nonzero angle of attack are neglected in this control model. However, it is possible to approximate these effects as first-order drag forces, as shown in Ref.~\cite{kai2017quad}.

\subsection{Aerodynamic forces}\label{sec-aero}

The resultant aerodynamic force $\bm F_a$ is traditionally decomposed into the sum of a drag and a lift force or resolved into components along the body frame as follows:
\begin{align} \label{eq:fa-1}
\begin{split}
\bm F_a &= \bm F_X + \bm F_Y+\bm F_Z \\
\bm F_X &=-\frac{1}{2} \rho S |\bm v_a|^2 C_X ~ \bm \imath \\
\bm F_Y &=-\frac{1}{2} \rho S |\bm v_a|^2 C_Y ~ \bm \jmath \\
\bm F_Z &=-\frac{1}{2} \rho S |\bm v_a|^2 C_Z ~ \bm k
\end{split}
\end{align}
where the aerodynamic coefficients $C_X$, $ C_Y$, and $ C_Z$ depend on the angle-of-attack $\alpha=atan({\frac{v_{a,3}}{v_{a,1}}})$, and the side-slip angle $\beta=asin({\frac{v_{a,2}}{|\bm v_a|}})$, as well as on the Reynolds and Mach numbers. In this work, we consider that the flight conditions are limited to low altitude and low airspeeds for which compressibility effects can be neglected; this allows us to reduce the dependency of the aerodynamic coefficients to $\alpha$ and $\beta$. We also note that we would like to constitute a control model, which allows the design of a cascaded modular control architecture as will be seen in later sections. This motivates the modeling of a triangular system where external control loops do not depend on higher order states. In the case of aerial vehicles, this means that the dependency of $\bm F_a$ on the angular velocities and control surfaces deflections should be neglected at this stage \footnote[1]{Unmodeled effects are compensated by feedback control loops, complemented with integral action. Subsequent simulations, with detailed and more representative models, are used to validate the applicability of the control design.}.
\begin{figure}[hbt!]
\centering
\includegraphics[width=.4\textwidth]{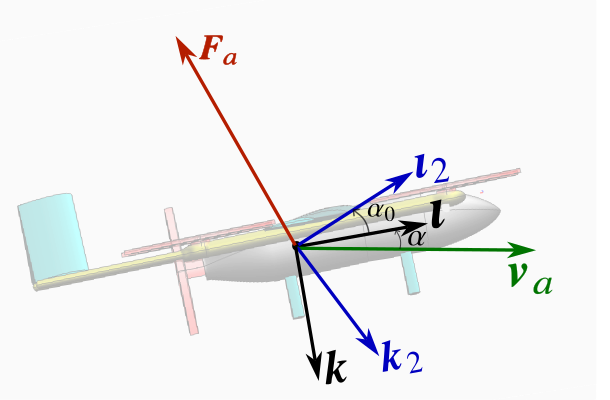}
\caption{Modified body-axis.}
\label{fig:modified-body-axis}
\end{figure}

Let ${\mathcal B_2}=\{G, \bm \imath_2, \bm \jmath, \bm k_2  \}$ represent a modified body-fixed frame, with $\bm \imath_2$ aligned with the zero-lift line of the vehicle, in other words, no lift force is generated if the relative air-velocity vector is aligned with $\bm \imath_2$. We consider that $\bm \imath_2$ is obtained by rotating $\bm \imath$ with an angle $\alpha_0$ around $\bm \jmath$ as shown in Fig.~\ref{fig:modified-body-axis}; this gives

\begin{equation}\label{eq:modified-body-axis}
\begin{cases}
\bm \imath_2 = \cos(\alpha_0) \bm \imath - \sin(\alpha_0) \bm k \\
\bm k_2 = \sin(\alpha_0) \bm \imath + \cos(\alpha_0) \bm k
\end{cases},~~~~
\begin{cases}
\bm \imath = \cos(\alpha_0) \bm \imath_2 + \sin(\alpha_0) \bm k_2 \\
\bm k = -\sin(\alpha_0) \bm \imath_2 + \cos(\alpha_0) \bm k_2
\end{cases}
\end{equation}
where $\alpha_0$ can be defined as the angle between the body-fixed vector $\bm \imath$ and the zero-lift line.

Considering the previous definitions and hypothesis, we propose to work with the more specific model of aerodynamic forces previously used in Refs.~\cite{2019-unified-automatica,2018thesis,2019soaring} and modified here to account for the zero-lift line,
\begin{equation}\label{eq:fa-2}
\bm F_a= -\frac{1}{2} \rho S |\bm v_a| (c_0 (\bm v_a \cdot \bm \imath_2) \bm \imath_2 + \barbar{c}_0 (\bm v_a \cdot \bm \jmath) \bm \jmath+\bar{c}_0 (\bm v_a \cdot \bm k_2) \bm k_2)
\end{equation}
With $c_0$, $\bar{c}_0$, and $\barbar{c}_0$ denoting positive coefficients. As stated in Refs.~\cite{2018thesis,2019-unified-automatica}, this model is compatible with Eq.~\eqref{eq:fa-1}, and corresponds to using $C_X=c_0 \cos(\alpha+\alpha_0) \cos(\beta)$, $C_Y=\barbar{c}_0 \sin(\beta)$ and $C_Z=\bar{c}_0 \sin(\alpha+\alpha_0) \cos(\beta)$.  These expressions represent bounded non-linear aerodynamic coefficients that do not grow unbounded when the angles $\alpha$ and $\beta$ get large; hence, they remain physically relevant and sufficiently representative for control design purposes by allowing analytical dynamic inversion in the inertial frame as will be seen in later section. Indeed, when the side-slip angle is equal to 0, the equivalent lift and drag coefficients are $C_L(\alpha)=\frac{1}{2}(\bar{c}_0- c_0)\sin(2 (\alpha + \alpha_0))$ and $C_D(\alpha)=c_0+(\bar{c}_0- c_0)\sin^2(\alpha + \alpha_0).$ When linearizing these expressions with respect to $\alpha$ and $\beta$, one can also deduce the following matches between coefficients: $c_0=C_{D,0}$, ~$\bar{c}_0=C_{D,0}+C_{L,\alpha}$,~ and $\barbar{c}_0=C_{Y,\beta}$~. The aerodynamic model in Eq.~\ref{eq:fa-2} is not necessarily representative beyond the stall region at high angle of attacks. However, when applying this model, the magnitude of the total aerodynamic force that results remains bounded at high airspeed, and it falls towards 0 when the airspeed approaches 0 where it remains well defined because it does not rely explicitely on $\alpha$ or $\beta$. This is of course a control model on the basis of which analytical flight control expressions will be derived. It remains to design reference trajectories which take into account the flight envelope limitations. For instance, in the case of the transition maneuver that goes from zero to cruising speeds, care must be taken to maintain the trimmed angle of attack below the stall region at high airspeed, and only allow higher angles of attack when the airspeed is low enough, and the vehile is in a proper hovering (multicopter) configuration.\footnote[4]{These aspects are taken into account in this paper when designing the transition strategy in Sec.~\ref{sec:trans-strat}.}

\subsection{Torque modeling}
\label{torque-mdl}

The aerodynamic torque $M_a$ is generally expressed in terms of nondimensional coefficients (static and stability derivatives) which are themselves functions of $\alpha$, $\beta$, and the components of the angular velocity $\omega$. For the sake of simplicity, we do not expand\footnote[2]{The level of uncertainty regarding these aerodynamic torque coefficients can sometimes be significant; this should not exclude the possibility to design flight control laws, especially if the design does not rely explicitly on these coefficients, as is the case in this paper.} this torque term and refer interested readers to the literature on the subject, for example, Refs.~\cite{nelson1998flight,aircraft-control}.
The control torque $M$ can be considered by itself a control input term. This allows the model and control design to be generalized to a large class of configurations. However, we will further develop the expression of $M$ for the compound VTOL configuration, as this will constitute our use case in subsequent simulations.

\begin{figure}[hbt!]
\centering
\includegraphics[width=.6\textwidth]{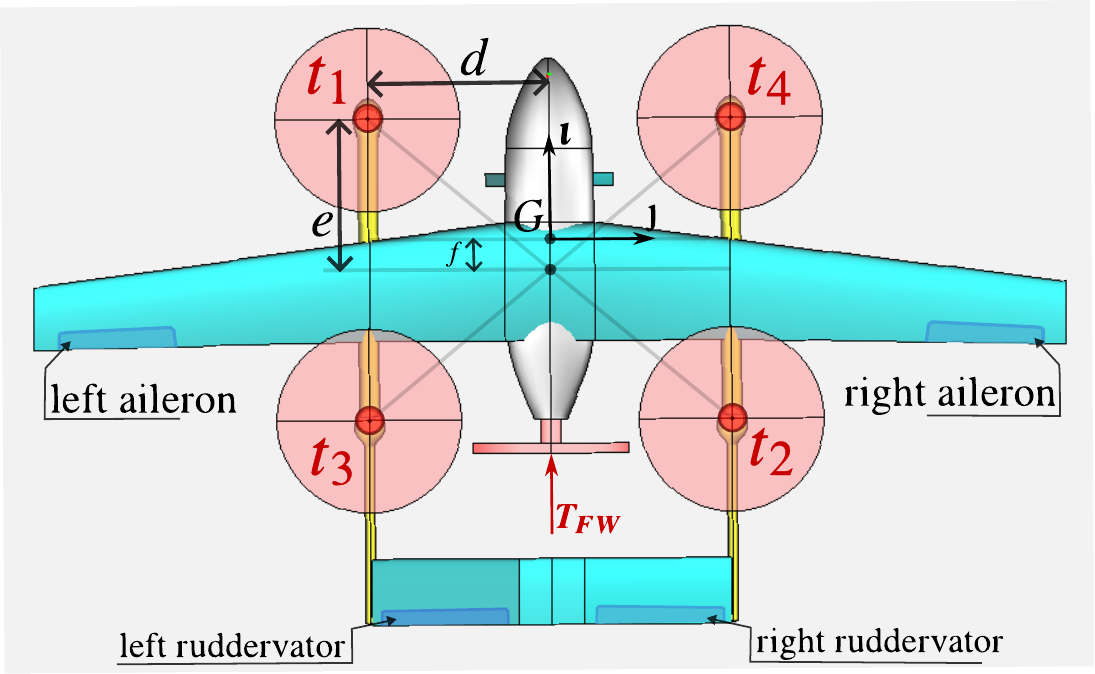}
\caption{Acturators configuration for a compound VTOL.}
\label{fig:actuators-config}
\end{figure}

Let $M_{MC}$ represent the torque generated by applying differential thrust on the lift rotors and $M_{FW}$ represent the torque generated by aerodynamic control surfaces. Therefore, we have $M=M_{MC}+M_{FW}$.
It is common to combine the torque vector $M_{MC}$ with the collective thrust generated by the lift rotors $|\bm T_{MC}|$ in one vector, that relates to the individual thrust forces $t_i$ generated by each rotor. We give the example here for a compound VTOL equipped with four lifting rotors:

\begin{equation}\label{eq:alloc-mc}
\begin{bmatrix}
|\bm T_{MC}|\\
M_{MC}
\end{bmatrix}
=
A
\begin{bmatrix}
t_1\\
t_2\\
t_3\\
t_4
\end{bmatrix}
, ~~~~~A=
\begin{bmatrix}
1 & 1 & 1 & 1 \\
d & -d & d & -d \\
e-f & -e-f& -e-f & e-f \\
\eta & \eta & -\eta & -\eta 
\end{bmatrix}
\end{equation}
where $d$, $e$ and $f$ are positive scalars characterizing the positions of the rotors with respect to the center of mass as shown in Fig.~\ref{fig:actuators-config}, and $\eta$ is a positive constant approximating the ratio of the power coefficient to the thrust coefficient of the lifting propellers near the hover condition.

Let $M_{FW}$ represent the total aerodynamic torque vector associated with the control surfaces. The vehicle we are considering is equipped with ailerons and an inversed V-tail consisting of two ruddervators. The former generates principally a roll torque, while the latter generate a pitching torque when deflecting in the same direction and a yawing torque when deflecting in opposite directions. A possible expression for  $M_{FW}$ as a function of the ailerons deflection\footnote[3]{The two ailerons are considered to be deflecting with the same angle in opposite directions; here, we adopt the convention of a positive angle $\delta_a$ corresponding to a positive rolling torque.} $\delta_a$ and the left and right ruddervators deflection $\delta_{rel}$ and $\delta_{rer}$ is the following:

\begin{equation}\label{eq:alloc-fw}
M_{FW}=\rho |\bm v_a|^2 B 
\begin{bmatrix}
\delta_a\\
\delta_{rel}\\
\delta_{rer}
\end{bmatrix}
,~~~~~B=\frac{1}{2} S
\begin{bmatrix}
b C_{l,\delta_a} & b C_{l,\delta_{rel}} & b C_{l,\delta_{rer}} \\
c C_{m,\delta_a} & c C_{m,\delta_{rel}} & c C_{m,\delta_{rer}} \\ 
b C_{n,\delta_{a}} & b C_{n,\delta_{rel}} & b C_{n,\delta_{rer}} \\ 
\end{bmatrix}
\end{equation}
with the control allocation matrix $B$ being a function of control derivatives that are commonly considered to be constants. In nominal conditions, the symmetrical functioning of the ruddervators leads to the following equalities between the coefficients: $C_{m,\delta_{rer}}=C_{m,\delta_{rel}}$ and $C_{n,\delta_{rer}}=-C_{n,\delta_{rel}}$.

\section{Control laws design}\label{sec:control}

We start by giving a simplified overview of the control design methodology based on the previously defined control model. Consider first that a feedback control function $\bm a_r$ is designed in order to achieve an objective related to the trajectory of the vehicle, such that if the translation dynamics equation is $\dot{\bm v}(t)= \bm a_r(\bm v(t), \bm v_r(t))$, the velocity $\bm v(t)$ will converge to a desired velocity $\bm v_r(t)$. The next step consists in solving the required attitude and thrust vector such that the sum of all the forces acting on the vehicule, including aerodynamic forces, results in an acceleration of the vehicle that matches the previously defined control term $\bm a_r$. In other words, we solve equation ${\bm a_r} = {\bm g} + \frac{{\bm F}_a}{m} + \frac{\bm T}{m} $  for the desired attitude of the vehicle represented by ${\mathcal B_r}=\{G, \bm \imath_r, \bm \jmath_r, \bm k_r  \}$ and the desired thrust vector represented by $(|\bm T_r|,\gamma_{T,r})$, such that in view of Eq.~(\ref{eq:trans-dyn}), the resulting translation dynamics would be $\dot{\bm v}(t)= \bm a_r$ . Finally, we design an attitude control function that makes the aircraft frame ${\mathcal B}=\{G, \bm \imath_, \bm \jmath, \bm k  \}$ converge to the previously defined desired attitude.

\begin{figure}[hbt!]
\centering
\includegraphics[width=1.0\textwidth]{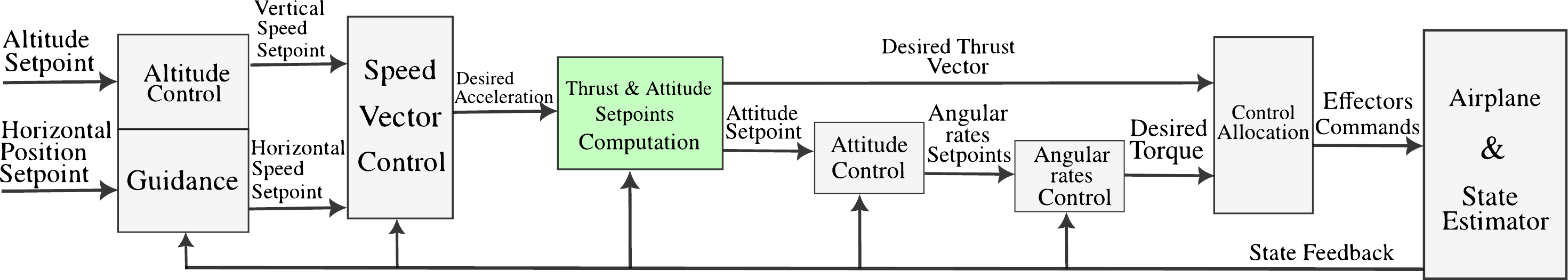}
\caption{Control Architecture.}
\label{fig:cont-arch}
\end{figure}

By following this control design methodology, we have adopted a hierarchical control architecture with outer position and speed control loops, and inner attitude and angular rate control loops as shown in Fig.~\ref{fig:cont-arch}. A key concept to keep in mind when designing such cascaded loops is the time scale separation, which means that each loop should have relatively faster convergence properties than its outer loop especially when feedforward terms are not transmitted on top of feedback terms. This requirement can be satisfied in the case of aerial vehicles because the inner attitude control loops are fully actuated and can be designed to have higher bandwidth with respect to the underactuated outer position/speed control loops.

Furthermore, hierarchical control offers many advantages by being modular and allowing us to use the same control modules for different piloting modes. For instance, the same modules can be applied for a fully autonomous flight or in the case of a pilot directly commanding speed vector components, altitude setpoints, attitude angles,  angular rates, thrust setpoints, or even combinations of the previous commands. It is also worth mentioning the practical benefit linked with facilitating the tuning and failure diagnosis procedures when working with separated lower order systems.
In the following subsections, each control module will be detailed by explaining its objective and the associated control law. These control laws build upon previous works in Ref.~\cite{2019-unified-automatica} in which they were applied to fixed-wing configurations with associated convergence and stability analysis, and then in Ref.~\cite{2019convertible} with an attempt to extend their applicability to convertible VTOL vehicles. Here, we adapt these laws to compound VTOL configurations and modify some control loops, in particular the \textit{thrust and attitude setpoint computation} step, which will be designed to be applicable in a unique way across the entire flight envelope, while being compatible with the chosen transition strategy.

\subsection{Position Control}

In this section, we detail the kinematical guidance loops where desired components of the speed vector $\bm v$ are computed to allow us to achieve an objective related to the positioning of the center of mass of the vehicle. 
 Two main types of position setpoints can be encountered:
\begin{itemize}
\item Trajectory tracking: the objective is to track a reference time-varying point in space. This autonomous mode is common for multi-copter vehicles with a hovering flight condition (zero speed) included in their flight domain.
\item Path following: the objective is to follow a path without time constraints. The center of mass of the vehicle should approach the path and follow it at a specified speed. This autonomous mode corresponds, for example, to classical waypoint following operations. It is also preferred for fixed-wing vehicles that apply a specified thrust and/or regulate their airspeed while adjusting their course to follow a desired path.
\end{itemize} 

We choose to separate the problem between a vertical altitude control and a horizontal guidance control. This choice is motivated by the transition strategy that we are aiming for and by the advantage of being able to maintain a unique altitude control loop functioning continuously all along the transition phase, even when the horizontal guidance switches from a trajectory tracking objective to a path following objective or vice versa.
\subsubsection{Altitude Control}
Let $z_r(t)$ denote the desired vertical position\footnote[4]{With a North-East-Down convention for the inertial frame, the reference altitude would be $-z_r$.}~. The objective of the altitude control module is to define a desired vertical speed $v_{z,r}(t)$ as an intermediate control variable that allows the convergence of the vertical position $z(t)= \bm p(t) \cdot \bm k_0$~ to ~$z_r(t)$, or equivalently the convergence of the tracking error $\tilde{z}=z-z_r$ to 0. We propose the following saturated proportional controller:
\begin{equation}\label{eq:cont-alt}
v_{z,r}= sat1D_{v_{z,min}}^{v_{z,max}}(-k_z \tilde{z}+\dot{z}_r)
\end{equation}
where the first feedback term is complemented with the feedforward term\footnote[5]{In some cases, when the vehicle is being piloted by a human, the feedforward term $\dot{z}_r$ is unknown and can be omitted. The convergence properties will then be limited by the bandwidth of the system.} $\dot{z}_r$, which accounts for time-varying reference inputs. The feedback gain $k_z$ is a positive scalar that determines the rate of convergence. The saturation limits $v_{z,min}$ and $v_{z,max}$ are chosen according to the flight envelope specifications of the vehicle and can take different values for multiple flight phases.
\subsubsection{Guidance: Trajectory Tracking}
Let $\bm r_{hor,r}(t) \in (\bm \imath_0, \bm \jmath_0)$ represent the reference position or trajectory of the vehicle in the horizontal plane. The objective of this module is to design an expression for the desired horizontal speed vector $\bm v_{hor,r}$ that ensures the convergence of the tracking error $\tilde{\bm r}_{hor}= \Pi_{\bm k_0} \bm r -\bm r_{hor,r}$ to 0. We propose the following saturated Proportional controller:
\begin{equation}\label{eq:cont-guidance-tt}
\bm v_{hor,r}=sat^{v_{h,max}}(-k_p \tilde{\bm r}_{hor} +\dot{\bm r}_{hor,r})
\end{equation}
Similarly to the altitude controller, we have a feedback term with a positive proportional gain $k_p$ complemented with a feedforward term. The desired speed is saturated at $v_{h,max}$.

\subsubsection{Guidance: Path Following}

Let $(\mathcal{C})$  be a pre-defined geometric path in the horizontal plane. We define the heading of the vehicle by the unitary vector $\bm h= \frac{\Pi_{\bm k_0} \bm{v}}{|\Pi_{\bm k_0} \bm{v}|}$.  The objective of this module is to compute a desired heading of the vehicle denoted by $\bm h_{r} \in (\bm \imath_0, \bm \jmath_0)$ that ensures the convergence to 0 of the distance between the aircraft's center of mass and $(\mathcal{C})$. Such approaches called \textit{vector field strategies} are well elaborated in the literature. We do not provide a particular solution here because the objective of a path following will not be necessary for our VTOL transition demonstration, and we refer interested readers to Refs.~\cite{2019-unified-automatica,vector-field}. We assume here that a desired heading $\bm h_{r}$ (mainly constant) will be defined, and bypasses this guidance module if necessary to be treated directly in the following \textit{speed vector control} module.

\subsection{Speed Vector Control}

The inputs for this module take the form of a desired vertical speed $v_{z,r}$ and a horizontal desired velocity expressed as a vector $\bm v_{hor,r}$ or as separated horizontal heading reference $\bm h_r$ and desired horizontal speed $v_{h,r} \in \mathbb{R}^+$. These inputs can either come from the outer \textit{position control} loops, or the outer loops can be bypassed and the inputs set directly here. Two types of horizontal velocity setpoints are identified; this leads to two separated horizontal speed control modules as shown in the following sections, with one of them being activated according to the type of the input.
The output of this module is a desired acceleration vector that will serve for the computation of the desired attitude and thrust vector of the vehicle. Therefore, the computation of the desired acceleration is to be viewed as an intermediate step, while the actual intermediate control terms are the attitude and the thrust vector.

\subsubsection{Vertical speed Control}

The objective of this module is to achieve the convergence of the vertical speed $v_z= \bm v\cdot \bm k_0$ to the setpoint $v_{z,r}$ by determining an intermediate desired vertical acceleration $a_{z,r}$. We define the tracking error $\tilde{v}_z=v_z-v_{z,r}$ and propose the following Proportional/Integral (PI) function:

\begin{equation} \label{eq:vert-speed-cont}
\begin{cases}
a_{z,r}= sat1D_{a_{z,min}}^{a_{z,max}}(-k_{vz} \tilde{v}_z- I_{vz}+\dot{v}_{z,r}) \\
\text{with,}\\
\frac{d}{dt} I_{vz} =
\begin{cases}
0, ~\textrm{if}~  (|I_{vz}| \geq \Delta_{I,vz}) \textrm{and}   (I_{vz}\tilde{v}_z > 0 )  \\
k_{I,vz} \tilde{v}_z ~\textrm{otherwise} 
\end{cases}
\end{cases}
\end{equation}
where $k_{vz}$, and $k_{I,vz}$ are positive proportional and integral gains. $I_{vz}$ is the integral term with its saturation value $\Delta_{I,vz} \in \mathbb{R}^+$. This saturation is taken into account to limit anti-windup effects\footnote[1]{For additional anti-windup protection, it is possible to limit integral action when the computed desired acceleration reaches the saturation values $a_{z,min}$ and $a_{z,max}$ .}. We also note that we start considering integral terms when dealing with dynamics that are subject to modeling uncertainties. Indeed, integral terms are known to add robustness properties to the control especially with respect to slowly time-varying disturbances. Note also that $a_{z,min}$ and $a_{z,max}$ are chosen to respect the flight envelope of the vehicle.

\subsubsection{Horizontal speed Control: Velocity Tracking}

The objective of this module is to track a horizontal time-varying velocity vector $\bm v_{hor,r}(t)$, by determining an intermediate desired horizontal acceleration $\bm a_{hor,r}(t) \in (\bm \imath_0,\bm \jmath_0)$. We define the tracking error by $\tilde{\bm v}_{hor}= \Pi_{\bm k_0} \bm v -\bm v_{hor,r}$, and propose the following PI function for $\bm a_{hor,r}$~:

\begin{equation}\label{eq:speed-control-tt}
\begin{cases}
\bm a_{hor,r}= sat^{a_{h,max}}(-k_{vh}\tilde{\bm v}_{hor}-  \bm I_{vh}+\dot{\bm v}_{hor,r}) \\
\text{with,}\\
\frac{d}{dt}  \bm I_{vh} =
\begin{cases}
\bm 0, ~\textrm{if}~  (|\bm I_{vh}| \geq \Delta_{I,vh}) \textrm{and}  (\bm I_{vh} \cdot \tilde{\bm v}_{hor}  > 0 )  \\
k_{I,vh} \tilde{\bm v}_{hor} ~\textrm{otherwise} 
\end{cases}
\end{cases}
\end{equation}
where $k_{vh}$ and $k_{I,vh}$ are positive proportional and integral gains, $\bm I_{vh}$ is the integral term, and $\Delta_{I,vh}  \in \mathbb{R}^+$ its associated saturation value. Note that  $a_{h,max}$ is the maximum horizontal acceleration value, which indirectly limits the bank angle of the vehicle. For multicopter platforms, and in the absence of aerodynamic forces, it can be shown that the maximum commanded bank angle will correspond to $atan(\frac{a_{h,max}}{g_0+a_{z,min}})$~.

\subsubsection{Horizontal speed Control:  Heading Tracking and Speed Regulation}

In this control mode, the horizontal speed control has two objectives:
\begin{itemize}
\item Speed regulation: convergence of the  aircraft horizontal speed $|\Pi_{\bm k_0} \bm v|$ to a setpoint $v_{h,r} \in \mathbb{R}^+$.
\item Heading tracking: convergence of the heading $\bm h= \frac{\Pi_{\bm k_0} \bm{v}}{|\Pi_{\bm k_0} \bm{v}|}$ to the desired heading $\bm h_{r} \in (\bm \imath_0, \bm \jmath_0)$.
\end{itemize}

By definition, we have the following expression for the horizontal acceleration of the aircraft: 
\begin{equation}
\begin{cases}\label{eq:def-a-sph}
\bm a_{hor} &= \frac{d}{dt} (\Pi_{\bm k_0} \bm{v}) \\
			&= \frac{d}{dt}(|\Pi_{\bm k_0} \bm v| \bm h) \\
          		&= \frac{d}{dt} (|\Pi_{\bm k_0} \bm v|) \bm h +|\Pi_{\bm k_0} \bm v| (\bm \omega_h \times \bm h)
\end{cases}
\end{equation}
where $\bm \omega_h = \bm h \times \dot{\bm h}$ denotes the angular velocity of the unitary heading vector $\bm h$. On the basis of this expression, we suggest dividing the desired horizontal acceleration $\bm a_{hor,r}$ into a tangential acceleration  $\bm a_{tan,r}$ parallel to $\bm h$ and a second lateral acceleration  $\bm a_{lat,r}$ orthogonal to the previous one. We define the speed regulation error as $e_v=|\Pi_{\bm k_0} \bm v| - v_{h,r}$ . An expression for $\bm a_{hor,r}$ is then given by the following:

\begin{equation}\label{eq:def-a-pf-cont}
\bm a_{hor,r} = \bm a_{tan,r} + \bm a_{lat,r} 
\end{equation}
with $\bm a_{tan,r}$ a saturated PI controller as
\begin{equation}\label{eq:def-atan-pf-cont}
\begin{cases}
\bm a_{tan,r}	 &= sat1D^{a_{t,max}}_{a_{t,min}}(-k_{t} e_v - I_{v,t} + \dot{v}_{h,r})  \bm h\\
\text{with,}\\
\frac{d}{dt}  I_{v,t} &=
\begin{cases}
0, ~\textrm{if}~  (|I_{v,t}| \geq \Delta_{I,vt}) \textrm{and}  (I_{v,t} e_v  > 0 )  \\
k_{I,t} e_v ~\textrm{otherwise} 
\end{cases}
\end{cases}
\end{equation}
where $k_t$, $k_{I,t}$ are positive gains, $\Delta_{I,vt}$ is the maximum value for the integrator, and $a_{t,min}$, $a_{t,max}$ are limitations for the tangential acceleration. In practice, and especially for fixed-wing platforms, it is more common to regulate the airspeed |$\bm v_a$| instead of the ground speed, by modifying the expression of the regulation error and using $e_v= |\bm v_a|- v_{h,r}$ instead. Indeed, maintaining the airspeed at nominal values helps reducing aerodynamic perturbations, preventing stall conditions, and ensuring an efficient flight condition for which the aircraft has been dimensioned.

A stabilizing feedback expression for $\bm a_{lat,r}$ is the following geometric PI controller (see Ref.~\cite{2019-unified-automatica}):

\begin{equation}\label{eq:def-alat-pf-cont}
\begin{cases}
\bm a_{lat,r} &= sat^{a_{l,max}}(|\Pi_{\bm k_0} \bm v| (\bm \omega_{h,r} \times \bm h)) \\
\text{with,}\\
\bm \omega_{h,r} &= k_h (\bm h \times \bm h_r) + \bm I_h + (\bm h_r \times \dot{\bm h}_r) \\
\frac{d}{dt}  \bm I_{h} &=
\begin{cases}
\bm 0, ~\textrm{if}~  (|\bm I_{h}| \geq \Delta_{I,h}) \textrm{and}  (\bm I_{h} \cdot (\bm h \times \bm h_r)  > 0 )  \\
k_{I,h} (\bm h \times \bm h_r) ~\textrm{otherwise} 
\end{cases}
\end{cases}
\end{equation}
where $k_h$, $k_{I,h}$ are positive parameters, $\Delta_{I,h}$ is the maximum value for the integrator, and $a_{l,max}$ is the maximum lateral acceleration. It can be shown that the maximum commanded bank angle will correspond approximately to $atan(\frac{a_{l,max}}{g+a_{z,min}})$~.

\subsubsection{Thrust Vector and Attitude Setpoints Computation}
\label{section-thrust-vectoring}

The outputs of the \textit{speed vector control} module can be summed in a single three-dimensional reference acceleration vector:
\begin{equation}\label{eq:desired-total-acc}
\bm a_r = a_{z,r} \bm k_0 + \bm a_{hor,r}
\end{equation}
This desired acceleration is a stabilizing feedback function for the outer loops. At this point, the objective is to impose this desired acceleration vector, through control terms defined by the thrust vector, and the attitude of the vehicle. Indeed, recalling the translation dynamics in Eq.~\eqref{eq:trans-dyn}, the objective of imposing $\dot{\bm v}=\bm a_r$ becomes equivalent to solving the following equation:

\begin{equation}
\bm a_r={\bm g} + \frac{{\bm F}_a}{m} + \frac{\bm T}{m}
\end{equation}
with $\bm F_a$ and $\bm T$ defined, respectively, in Eqs.~\eqref{eq:fa-2} and \eqref{eq:thrust-dir}. 
Let $\mathcal{B}_r=(\bm \imath_r, \bm \jmath_r, \bm k_r)$ represent the desired body frame and $(|\bm T_r|,\gamma_{T,r})$ represent the norm and direction of the desired thrust vector. We propose the solution from Ref.~\cite{2019convertible}, that is modified here to become applicable to hovering flights with $\bm v_a = \bm 0$. This is done by not relying directly on the angle of attack in the computation of the desired attitude, but on a vector $\bm a'$ that is always defined along the vehicle's trajectory. Indeed, $\alpha$ becomes ill-defined when the air-velocity $\bm{v}_a$ nears $\bm{0}$. In addition, the proposed solution takes into account a misalignement of the zero lift axis with the forward axis of the vehicle $\bm \imath$.~A detailed proof is reported in appendix \ref{sec:app1}. 

 We first define the following:
\begin{align}
\bm a' &= \bm a_r - \bm g \label{eq:th-sol1}\\
\bm d &= m \bm a' + \frac{1}{2} \rho S |\bm v_a|c_0 \bm v_a \label{eq:th-d} \\
\bm e &= m \bm a' + \frac{1}{2} \rho S |\bm v_a| \bar{c}_0 \bm v_a \label{eq:th-e} 
\end{align}

The expression of $\bm \jmath_r$ is defined based on one of the two objectives:
\begin{itemize}
\item Imposing a desired yaw angle $\psi_r$, in which case we propose the following expression for $\bm \jmath_r$:
\begin{equation}\label{eq:jr1}
\begin{cases}
\bm \jmath_r & =\frac{\bm h_{\psi,r} \times \bm a'}{|\bm h_{\psi,r} \times \bm a'|} \\
\bm h_{\psi,r} &=[\cos(\psi_r), \sin(\psi_r), 0]^\top
\end{cases}
\end{equation}
\item Zeroing the side-slip angle $\beta$, which corresponds to adopting a \textit{balanced flight} or a \textit{coordinated turn}. This is also equivalent to having the airspeed vector and the aerodynamic forces in the longitudinal plane of the aircraft $(\bm \imath, \bm k)$. Therefore, $\bm \jmath_r$ should be orthogonal to both the air-velocity $\bm v_a$ and the apparent acceleration $\bm a'$:
\begin{equation}\label{eq:jr2}
\bm \jmath_r = \frac{\bm v_a \times \bm a'}{|\bm v_a \times \bm a'|}
\end{equation}

The first objective of tracking a yaw angle is mostly employed when piloting multicopter platforms. To avoid the singularity at $|\bm h_{\psi,r} \times \bm a'|=0$, one can choose a value of $a_{z,max}$ (maximum downward acceleration) inferior to the gravity acceleration $g$ in Eq.~\eqref{eq:vert-speed-cont}.
The second objective of a balanced flight is employed in the case of fixed-wing platforms that fly at speeds that are sufficient enough to allow the estimation of the air-velocity vector $\bm v_a$. As discussed in Ref.~\cite{2019-unified-automatica}, the condition $|\bm v_a \times \bm a'|\neq 0$ is not restrictive in practice and includes a much larger set than classically defined trim trajectories. In practice, divisions by 0 are avoided \footnote[2]{A possibility is to zero out the vectors at the singularity, which later leads to a null desired angular velocity vector. If this situation is reached, it does not exclude the convergence again to the equilibirum of the system because a small perturbation is sufficient to get the system out of the singularity.}~. 
\end{itemize}

We exploit here the thrust-vectoring capabilities of VTOL platforms, in two ways, either by imposing the thrust direction $\gamma_T$ (as in Ref.~\cite{2019convertible}) or by imposing the equivalent of a pitch angle for the vehicle attitude. We give next the expressions for the control terms that result from the proofs reported in the Appendix.

\begin{itemize}
\item \textbf{case 1:} Imposed thrust direction

Here, the value of $\gamma_T$ is imposed at $\gamma_{T,r}$. The expressions for the remaining desired body frame axis are the following:

\begin{equation}\label{cont-imp-thrust-dir}
\begin{cases}
\bm k_r & = \sin(\gamma) \frac{\bm a'}{|\bm a'|} + \cos(\gamma) \frac{\bm a^{'\perp}}{|\bm a^{'\perp}|}  \\
\bm \imath_r &= \bm \jmath_r \times \bm k_r\\
\text{with,}\\
\bm a^{'\perp} &= \bm a' \times \bm \jmath_r \\
y &= \sin(\gamma_{T,r} + \alpha_0) \bm d \cdot \bm a' - \cos(\gamma_{T,r} + \alpha_0) \bm e \cdot \bm a^{'\perp} \\
x&= \cos(\gamma_{T,r} + \alpha_0) \bm e \cdot \bm a' + \sin(\gamma_{T,r} + \alpha_0) \bm d \cdot \bm a^{'\perp}\\
\gamma &= atan2(y,x) - \alpha_0 

\end{cases}
\end{equation}

\item \textbf{case 2:} Imposed pitch angle

Here a desired pitch angle denoted by $\theta_r$ is imposed. The expressions for the remaining desired body frame axis and the desired thrust direction are the following:

\begin{equation}\label{cont-imp-pitch}
\begin{cases}
\bm \imath_r & = \cos(\theta_r) \bm \eta + \sin(\theta_r) \bm \eta^{\perp} \\
\bm k_r &= \bm \imath_r \times \bm \jmath_r\\
\gamma_{T,r} &=atan2(y',x')-\alpha_0 \\
\text{with,}\\
\bm \eta &=\frac{\bm \jmath_r \times \bm k_0}{|\bm \jmath_r \times \bm k_0|} \\
\bm \eta^{\perp} &=\frac{\bm \jmath_r \times \bm \eta}{|\bm \jmath_r \times \bm \eta|} \\
y' &=  \sin(\alpha_0) \bm e \cdot \bm \imath_r + \cos(\alpha_0) \bm e \cdot \bm k_r \\
x' &= \cos(\alpha_0) \bm d \cdot \bm \imath_r - \sin(\alpha_0) \bm d \cdot \bm k_r

\end{cases}
\end{equation}

\end{itemize}

For both cases, the desired thrust norm is given by:
\begin{align}\label{cont-thr}
\begin{split}
|\bm T_r| =& \cos( \gamma_{T,r} + \alpha_0 ) \cos(\alpha_0) \bm d \cdot \bm \imath_r - \cos(\gamma_{T,r} + \alpha_0) \sin(\alpha_0) \bm d \cdot \bm k_r \\
&+ \sin(\gamma_{T,r} + \alpha_0) \sin(\alpha_0) \bm e \cdot \bm \imath_r + \sin(\gamma_{T,r} + \alpha_0) \cos(\alpha_0) \bm e \cdot \bm k_r
\end{split}
\end{align}

 In the case of compound configurations, one can deduce the values for the desired thrusts $|\bm T_{MC,r}|$ and $| \bm T_{FW,r}|$ from $|\bm T_r|$ and $\gamma_{T,r}$ using the expressions in Eq.~\eqref{eq:seperated-thrust}. Additionally, one can impose saturations on the computed thrust values according to the limitations of the rotors capabilities.  

 Convergence analysis for the translation dynamics with the preceding control term expressions is reported in Appendix \ref{sec:app12} and allows us to state the following theorem for the translation dynamics convergence.

\begin{theorem}
\label{th:trans-dyn-conv}
Consider a vehicle whose translation dynamics is governed by Eq.~\eqref{eq:trans-dyn}, where the applied thrust and aerodynamic forces are modeled accoding to Eqs.~\eqref{eq:trans-dyn} and \eqref{eq:fa-2} respectively. Given a desired acceleration vector $\bm a_r$, define the desired body-fixed frame  ${\mathcal B_r}=\{G, \bm \imath_r, \bm \jmath_r, \bm k_r \}$ and desired thrust vector $(|\bm{T}_r|,\gamma_{T,r})$ according to Eqs.~\mref{eq:th-sol1,cont-thr}. Assume that the thrust commands $\bm T_{MC,r} =  |\bm T_r| \sin(\gamma_{T,r})\bm k$ and $\bm T_{FW,r} =  |\bm T_r| \cos(\gamma_{T,r})\bm \imath$ are tracked by the propulsion systems and that $|\bm d|$, $|\bm e|$, and $|\bm a'+ \frac{1}{2}\rho S|\bm v_a| \barbar{c}_0 \bm v_a|$ are bounded. An attitude control law that achieves the convergence of the body-fixed frame ${\mathcal B}=\{G, \bm \imath, \bm \jmath, \bm k  \}$ to a well-defined desired frame ${\mathcal B_r}$ leads to the closed-loop translation dynamics of the form $\dot{\bm v}=\bm a_r + \bm o(t)$ with $\displaystyle{\lim_{t \to \infty}} \bm o(t)=0$.

\end{theorem}

Note that the first case of imposing the thrust direction reduces to conventional multicopter piloting when imposing $\gamma_{T,r}=-\frac{\pi}{2}$ and neglecting the aerodynamic forces (by imposing $c_0=0$ and $\bar{c}_0=0$). Indeed, we find again the classical geometric control that is widely used today in open-source autopilots but also in research projects such as Refs.~\cite{hamel2002dynamic,ackerman,mahonyquad,kai2017quad}. This same approach is extended to fixed-wing applications by including the aerodynamic forces in the control model; indeed, it suffices to fix  $\gamma_{T,r}=0$ to find the control proposed in Ref.~\cite{2019-unified-automatica}. The general solution presented here is generic, and it presents a common control design for different configurations and is therefore particularly suitable for VTOL transition maneuvers with all intermediate states encountered during vehicle reconfiguration.

	As stated in theorem \ref{th:trans-dyn-conv}, the translation dynamics convergence requires the design of an attitude control law which achieves the convergence of the aircraft frame ${\mathcal B}$ to a desired frame ${\mathcal B_r}$. This will be the objective of the following sections.

\subsection{Attitude Control}

The objective of this module is to make the aircraft frame ${\mathcal B}=\{G, \bm \imath, \bm \jmath, \bm k  \}$ converge to the desired frame ${\mathcal B_r}=\{G, \bm \imath_r, \bm \jmath_r, \bm k_r  \}$. This can be done using the following geometric control expression for the desired angular velocity in the inertial frame (see Ref.~\cite{2019-unified-automatica}):

\begin{equation}
\begin{cases}\label{eq:att-cont}
\bm \omega_r &= k_{\mathcal{B}} ( (\bm \imath \times \bm \imath_r) + (\bm \jmath \times \bm \jmath_r) + (\bm k  \times \bm k_r)) + \bm \omega_{ff}  \\
\text{with,}\\
\bm \omega_{ff} &= (\bm k_r \times \frac{d \bm k_r}{dt}) + ((\bm \jmath_r \times \frac{d \bm \jmath_r}{dt})  \cdot \bm k_r) \bm k_r
\end{cases}
\end{equation}
Where $k_{\mathcal{B}}$ is a positive proportional gain and $\bm \omega_{ff}$ is a feedforward term that corresponds to the angular velocity associated with the frame ${\mathcal B_r}$.

An alternative expression for the control term in Eq.~\eqref{eq:att-cont}, that allows to associate seperate gains with each body-frame axis is the following:

\begin{equation}
\begin{cases}\label{eq:att-cont2}
\bm \omega_r &= k_{\bm \imath}( \bm{\omega_0}\cdot \bm \imath) \bm \imath + k_{\bm \jmath}( \bm{\omega_0}\cdot \bm \jmath) \bm \jmath + k_{\bm k}(\bm{\omega_0}\cdot \bm k) \bm k + \bm \omega_{ff}\\
\text{with,}\\
\bm{\omega_0} &= (\bm \imath \times \bm \imath_r) + (\bm \jmath \times \bm \jmath_r) + (\bm k  \times \bm k_r) \\
\bm \omega_{ff} &= (\bm k_r \times \frac{d \bm k_r}{dt}) + ((\bm \jmath_r \times \frac{d \bm \jmath_r}{dt})  \cdot \bm k_r) \bm k_r
\end{cases}
\end{equation}

where $k_{\bm \imath}$, $k_{\bm \jmath}$ and $k_{\bm k}$ can be interpreted as respectively the roll, pitch and yaw proportional gains. A proof for asymptotic convergence associated with this last control term is reported in Appendix \ref{sec:app2}.

\subsection{Angular Rate Control}

The objective of this module is to make the aircraft's angular velocity $\bm \omega$ converge to the desired angular velocity vector $\bm \omega_r$. As shown in Eq.~\eqref{eq:rot-dyn}, the attitude dynamics are expressed in the body frame. Thus, we elaborate here the control functions using the vectors of coordinates $\omega$ and $\omega_r$ expressed in the body frame. We define the tracking error as $\tilde{\omega}=\omega-\omega_r$; then on the basis of the state equation  (Eq.~\eqref{eq:rot-dyn}), we propose the following PI function for a desired control torque $M_r$:

\begin{equation} \label{eq:ang-rate-cont}
\begin{cases}
M_r= - K_{P,\omega} J \tilde{\omega} - I_{\omega} + \omega \times J \omega - M_a + J \dot{\omega}_r \\
\text{with,}\\
\frac{d}{dt} I_{\omega,i} =
\begin{cases}
0, ~\textrm{if}~  (|I_{\omega,i}| \geq \Delta_{I,\omega_i}) \textrm{and}   (I_{\omega,i}\tilde{\omega}_i > 0 )  \\
k_{I,\omega_i} ~\tilde{\omega}_i  ~\textrm{otherwise}  \\
\textrm{for}~ i=1,2,3
\end{cases}
\end{cases}
\end{equation}

where $K_{P,\omega}=diag(k_{p,\omega_1},k_{p,\omega_2},k_{p,\omega_3})$ is a diagonal matrix composed of three positive proportional gains. And $k_{I,\omega_1}$, $k_{I,\omega_2}$ and $k_{I,\omega_3}$ are three positive integral gains. The positive scalars $\Delta_{I,\omega_i}$ define saturation values for the integrators.

It is possible in practice to omit the cancelation of the aerodynamic torque $M_a$. For most aircraft, this passive torque contributes to the stability of the attitude dynamics around trim trajectories; it is classically taken into account for the tuning of linear controllers\footnote[3]{Examples of such tuning techniques are pole placement, eigenstructure assignments, Linear Quadratic Regulators (LQR), and robust H$\infty$ methods. See, for example, Ref.~\cite{aircraft-control}.}, without necessarily being exactly compensated. Furthermore, for some UAV applications, the identification of stability derivatives can be challenging, and an estimation of the vector $M_a$ might not be available. However, this should not exclude the possibility to design control laws, especially for the attitude dynamics, which are fully actuated, and can in general tolerate high control gains. For all these reasons, and considering the relatively higher bandwidth of the attitude control loops, we can further neglect the Coriolis and feedforward terms and employ the following alternative expression for $M_r$:

\begin{equation}
\label{eq:ang-rate-cont2}
M_r= - K_{P,\omega} J \tilde{\omega} - I_{\omega}
\end{equation}

Note that this intermediate step of obtaining a control torque $M_r$ is still independent of the actuation configuration of the vehicle.

\subsection{Control Allocation Mapping}

From this point, we give the example of control allocation mapping to the effectors in the case of a compound vehicle as it was modelled in Sec.~\ref{torque-mdl}. The inputs of this module are the commanded thrust vector $\bm T_r$ divided into two components $\bm T_{MC,r}$ and $\bm T_{FW,r}$ and the control torque $M_r$. Recall that the values for the desired thrusts $|\bm T_{MC,r}|$ and $|\bm T_{FW,r}|$  can be deduced from s$|\bm T_r|$ and $\gamma_{T,r}$ using the expressions in Eq.~\eqref{eq:seperated-thrust} as follows:
\begin{align}
	|\bm T_{MC,r}| &= |\bm T_r| |sin(\gamma_{T,r})| \\
	|\bm T_{FW,r}| &= |\bm T_r| |cos(\gamma_{T,r})| 
\end{align}

We define the scalar $\lambda \in [0,1]$ as the blending coefficient for torque allocation. When $\lambda$ is set to $0$, the control torque should be generated entirely by the differential thrust of the lifting rotors. On the other hand, when $\lambda$ is set to $1$, the control torque is generated by the aerodynamic control surfaces. This allows us to define  the distribution of $M_r$ between a component $M_{MC,r}$ to be generated by the lifting rotors and a component $M_{FW,r}$ to be generated by the control surfaces. We choose the following continuous blending expressions with respect to $\lambda$:

\begin{equation}
\begin{cases}\label{eq:alloc-1}
M_{MC,r} &=(1-\lambda) M_r \\
M_{FW,r} &= \lambda M_r
\end{cases}
\end{equation}

The individual thrust values of each lifting rotor, can be computed in accordance with Eq.~\eqref{eq:alloc-mc}, via the inversion\footnote[4]{In general, for multicopter configurations with more than four rotors, the matrix $A$ would not be a square matrix, in which case one would consider using the pseudo-inverse operator $A^{+}=A^{\top} (A A^{\top})^{-1}$.} of the matrix $A$, in other words,

\begin{equation}\label{eq:alloc-mc-inv}
\begin{bmatrix}
t_{1,r}\\
t_{2,r}\\
t_{3,r}\\
t_{4,r}
\end{bmatrix}
=
A^{-1}
\begin{bmatrix}
|\bm T_{MC,r}|\\
M_{MC,r}
\end{bmatrix}
\end{equation}

The deflection angles of control surfaces are computed in accordance with Eq.~\eqref{eq:alloc-fw}, as follows:

\begin{equation}\label{eq:alloc-fw-inv}
\begin{bmatrix}
\delta_{a,r}\\
\delta_{rel,r}\\
\delta_{rer,r}
\end{bmatrix}
=\frac{1}{\rho |\bm v_a|^2}B^{-1} M_{FW,r}
\end{equation}

In practice, desaturation strategies are implemented in addition to the previous allocation mapping in order to deal with situations where certain components of the commanded thrusts or control surfaces angles are outside of their feasible range. We do not detail a specific desaturation strategy in this paper for the sake of brevity and refer the interested reader to the abundant literature on the subject. For instance, the most common algorithms adopt a prioritized policy that favors pitch and roll torque realizations (see  Refs.~\cite{alloc1,alloc2}).

\section{High-Level Transition Strategy}\label{sec:trans-strat}

The transition strategy that we propose here takes the form of successive high-level setpoints, that progressively achieve the reconfiguration of the vehicle from a hovering flight to a high-speed cruise flight and inversely. In this sense, we are exploiting the applicability of the previously designed control loops to the entire flight envelope and restricting the design activity at this stage to the establishment of an adequate design of transition flight phases.

\begin{figure}[hbt!]
\centering
\includegraphics[width=0.75\textwidth]{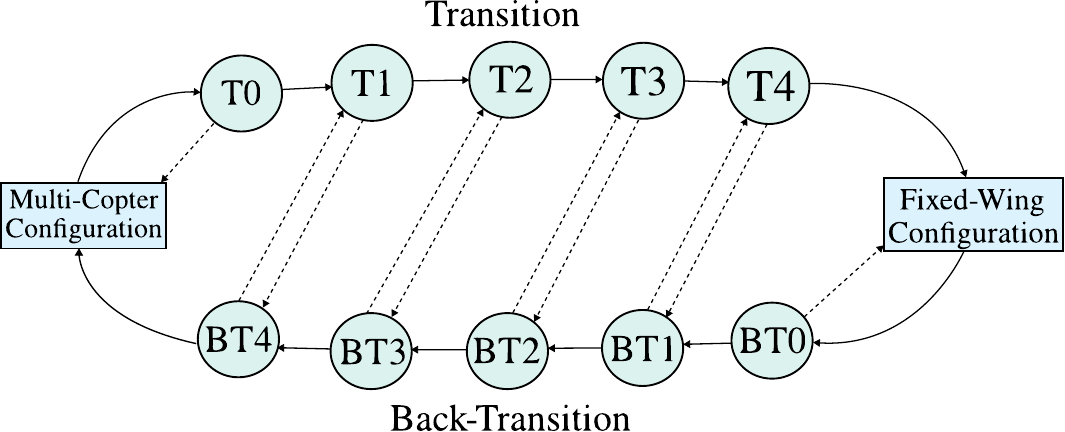}
\caption{Transition and Back-Transition State Machine.}
\label{fig:state-machine}
\end{figure}

The sequence of the flight phases are represented in the form of the state machine shown in the Fig.~\ref{fig:state-machine}. We assume that the vehicle takes off vertically and is piloted as a multicopter vehicle. The pilot or the operator may then specify the desired heading and initial altitude for the transition and command the transition profile to be executed according to the state machine with specific parameters that shape the trajectory profile. In a similar way, the pilot or operator can initiate the back-transition maneuver from a cruise flight at a proper location and altitude. Another asset of the proposed transition strategy, is the ability to abort the transition at any intermediate state by progressing towards the appropriate state of the back-transition maneuver. Indeed the transition and back-transition states are designed to be analogous. 

The forward conditions that clear the passage to the following state can simply take the form of convergence criteria to desired high-level setpoints. For example, one may specify an acceptable threshold $\epsilon_{\tilde{x}}>0$ for each tracking error $\tilde{x}$, in other words, $|\tilde{x}|<\epsilon_{\tilde{x}}$. The backward conditions represented by dashed lines in Fig.~\ref{fig:state-machine} basically abort the transition. This can result from a pilot abort command, a failure detection, a divergence condition, or even \textit{timeouts} suggesting an anomaly had occurred during the execution of the phase in question. Tables \ref{tab:table1} and \ref{tab:table2} describe each of the transition phases (T0 to T4) and the back-transition phases (BT0 and BT4) as well as the associated high-level setpoints.

\subsection{Transition flight phases}

\begin{table}[hbt!]
\caption{\label{tab:table1} Transition flight phases description}
\begin{tabular}{lll}
\hline
Phase& Description&  Setpoints\\\hline
MC& The vehicle is piloted as a multicopter. & $\gamma_{T,r}=-\frac{\pi}{2}$, $\lambda=0$\\\hline
T0& Beginning of the transition from a hovering state, a pitch setpoint is specified, & $\theta_{T0}$, $\lambda=0$\\
    & with a desired climb rate, a desired yaw,  & $v_{z,T0}<0$, $\psi_r$\\
    &and a desired horizontal speed vector increasing in magnitude. & $\bm v_{hor,r}$\\\hline
T1& A pitch setpoint is specified, & $\theta_{T1}$, $\lambda=0$\\
    & with a desired climb rate, sideslip cancellation, & $v_{z,T1}<0$, $\beta=0$\\
    & a desired airspeed, and a desired horizontal heading. & $v_{a,T1}$, $\bm h_{r}$\\\hline
T2& A pitch setpoint is specified, the torque control blended continuously toward& $\theta_{T2}$, $\lambda=0\longrightarrow 1 $\\
    &  control surfaces, with a desired climb rate, sideslip cancellation, & $v_{z,T2}<0$, $\beta=0$\\
    & a desired airspeed, and a desired horizontal heading. & $v_{a,T1}$, $\bm h_{r}$\\\hline
T3& A pitch setpoint is specified, & $\theta_{T3}$, $\lambda=1$\\
    & with a desired climb rate, sideslip cancellation, & $v_{z,T3}<0$, $\beta=0$\\
    & a desired airspeed increasing toward cruise speed, and a desired horizontal heading. & $v_{a,FW}$, $\bm h_{r}$\\\hline
T4& The vehicle is piloted in a fixed-wing configuration, & $\gamma_{T,r}=0$, $\lambda=1$\\
    & at a fixed altitude, with sideslip cancellation, & $z_r$, $\beta=0$\\
    & a fixed airspeed at nominal cruise speed, and a desired horizontal heading. & $v_{a,FW}$, $\bm h_{r}$\\\hline
FW& The vehicle is piloted as a fixed-wing aircraft. & $\gamma_{T,r}=0$, $\lambda=1$\\\hline
\hline
\end{tabular}
\end{table}

The transition begins with phase T0, where the pitch is imposed to a specific value, which will activate the pusher rotor on top of the lifting rotors in order to push the vehicle forward. A desired horizontal speed vector is specified with an increasing magnitude allowing us to reach airspeed values at the end of T0 that are sufficient enough to ensure that $\bm v_a \neq 0$ and that $\bm h$ is well defined. This allows us in phase T1 to track a desired heading of the vehicle and to adopt a balanced flight strategy (by zeroing the side-slip). At this point, the airspeed is accelerated until it reaches an intermediate value $v_{a,T1}$ for which the aerodynamic control surfaces can produce enough torque actuation for the controllability of the attitude dynamics. The airspeed value $v_{a,T1}$ should also be chosen so that enough collective thrust $T_{MC}$ is produced to avoid reducing differential thrust authority so that a control torque can still be produced simultaneously whether by lifting rotors or control surfaces. Indeed, in phase T2, we increase the blending factor $\lambda$ progressively from $0$ to $1$ during a fixed duration (typically 2 seconds), this will transfer the torque actuation entirely to control surfaces. In phase T3, the vehicle is accelerated toward cruise speed, with a fixed pitch angle $\theta_{T3}$ that is chosen here to be close to the trimming angle of attack at nominal cruise speed. During T3, it is expected that the aerodynamic lift force is increasing until it can compensate the weight of the vehicle so that the thrust generated by the lifting rotors is sharply reduced. Finally, in phase T4 the vehicle is controlled in a pure , fixed-wing configuration, with a stabilization of its altitude, heading, and airspeed.

\subsection{Back-Transition flight phases}

\begin{table}[hbt!]
\caption{\label{tab:table2} Back-Transition flight phases description}
\begin{tabular}{lll}
\hline
Phase& Description&  Setpoints\\\hline
FW& The vehicle is piloted as a fixed-wing aircraft. & $\gamma_{T,r}=0$, $\lambda=1$\\\hline
BT0& The vehicle is piloted in a fixed-wing configuration, & $\gamma_{T,r}=0$, $\lambda=1$\\
    & with a desired descent rate, with side-slip cancellation, & $v_{z,BT0}>0$, $\beta=0$\\
    & a fixed airspeed at nominal cruise speed, and a desired horizontal heading. & $v_{a,FW}$, $\bm h_{r}$\\\hline
BT1& A pitch setpoint is specified, & $\theta_{BT1}$, $\lambda=1$\\
    & with a desired descent rate, with sideslip cancellation, & $v_{z,BT1}>0$, $\beta=0$\\
    & a fixed airspeed at nominal cruise speed, and a desired horizontal heading. & $v_{a,FW}$, $\bm h_{r}$\\\hline
BT2& A pitch setpoint is specified, & $\theta_{BT1}$, $\lambda=1$\\
    & with a desired descent rate, with side-slip cancellation, & $v_{z,BT2}>0$, $\beta=0$\\
    & a deceleration towards an intermediate airspeed, and a desired horizontal heading. & $v_{a,BT2}$, $\bm h_{r}$\\\hline
BT3& A pitch setpoint is specified, the torque control blended continuously toward & $\theta_{BT3}$, $\lambda=1\longrightarrow 0 $\\
    & differential thrust, at a fixed altitude, with side-slip cancellation, & $z_{BT3}$, $\beta=0$\\
    & at a fixed intermediate airspeed, and a desired horizontal heading. & $v_{a,BT2}$, $\bm h_{r}$\\\hline
BT4& The vehicle is piloted in a multicopter configuration, & $\gamma_{T,r}=-\frac{\pi}{2}$, $\lambda=0 $\\
    & at a fixed altitude, with a desired yaw, & $z_{BT4}$, $\psi_r$\\
    & and a desired horizontal speed vector decreasing toward $\bm{0}$ & $\bm v_{hor,r}$\\\hline
MC& The vehicle is piloted as a multicopter. & $\gamma_{T,r}=-\frac{\pi}{2}$, $\lambda=0$\\\hline
\hline
\end{tabular}
\end{table}

The back-transition begins with phase BT0, which is essentially a descent phase allowing the aircraft to approach its landing site. In phase BT1, the pitch is imposed; this will activate the lifting rotors, which will allow the vehicle to decelerate its airspeed in BT2 while maintaining the descent rate control without stalling. Indeed, the vehicle reaches the intermediate speed $v_{a,BT2}$ which can be chosen equal to $v_{a,T1}$. In BT3, the blending factor is varied progressively from $1$ to $0$ during a fixed duration (typically less than 2 seconds); this will transmit the generation to torque control entirely to the lifting rotors. The altitude is locked and regulated at its value at the beginning of the phase. Finally, in phase BT4, side-slip control is abandoned to yaw control, and the vehicle is piloted as a multicopter with a decelerating horizontal speed towards 0.

Variants of the proposed transition and back-transition state machines can be proposed. For instance, climb and descent rates can be replaced everywhere with a fixed altitude setpoint in order to achieve 'horizontal' transition maneuvers. One can also choose to keep imposing $\gamma_{T,r}=-\frac{\pi}{2}$ in phase T0 so that the vehicle moves forward by inclining the collective thrust of the lifting rotors, instead of activating the pusher propeller at this early stage where its efficiency might be poor. To guarantee a narrow flight corridor, it is also possible to follow a straight horizontal path during the transition and back-transition instead of tracking a fixed heading. Finally, instead of executing the transition and back-transition phases in a complete autonomous mode, it is possible to leave to the pilot or  the remote operator the possibility to steer the vehicle in terms of heading, climb or descent rate and airspeed target.

\section{Implementation and Hardware-In-The-Loop Simulations}\label{sec:HIL}

\begin{figure}[hbt!]
\centering
\includegraphics[width=0.35\textwidth]{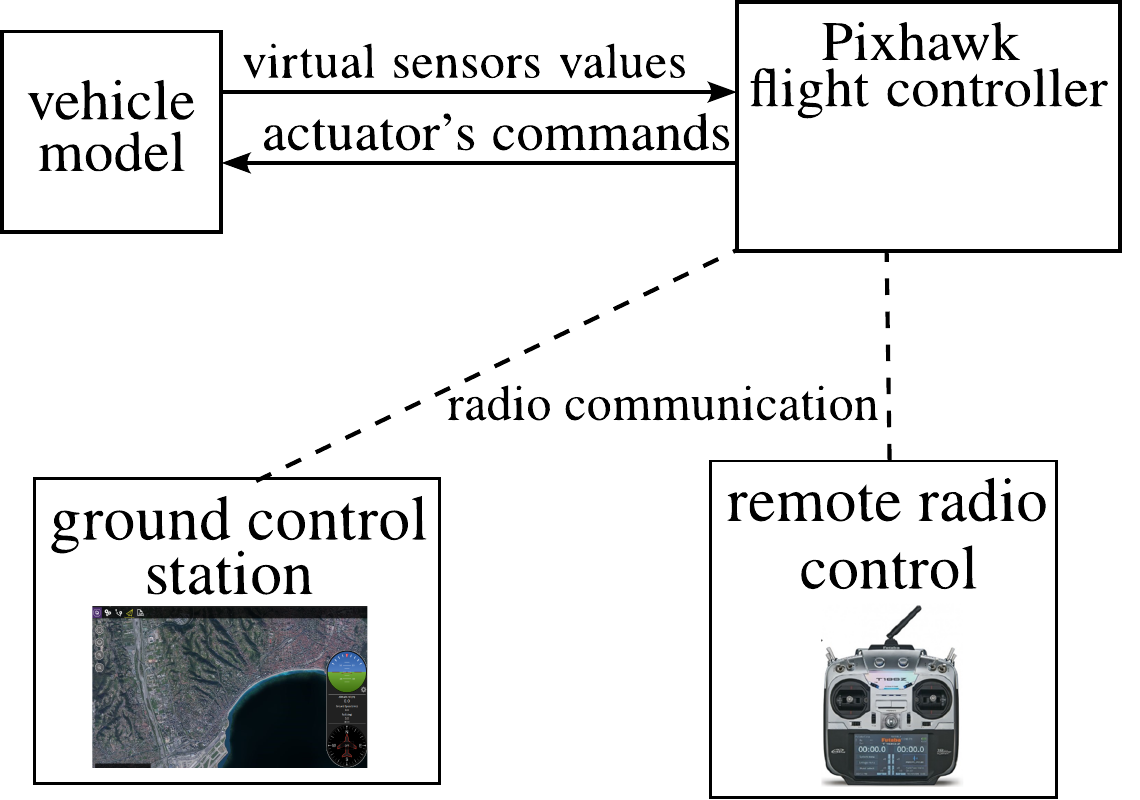}
\caption{HIL setup.}
\label{fig:HIL-setup}
\end{figure}

In this section, we test the proposed control laws and transition strategy in a Hardware-In-The-Loop (HIL) simulation. The HIL setup consists of a Matlab\textregistered/Simulink\textregistered~model of the vehicle connected to a commercially available "Pixhawk 3 Pro" autopilot flight controller, where virtual sensors values are sent to the controller, who in turn computes the actuator's commands and sends them back, as shown in Fig.~\ref{fig:HIL-setup}. The flight control implementation is based on the open-source PX4 flight stack, where the native control modules were replaced by the flight control algorithm and the transition state machine that are presented in this paper. Other functionalities of the PX4 autopilot are kept, in particular the Extended Kalman Filter (EKF) state observer.

The targeted vehicle is a commercially available compound VTOL, with a weight of $18Kg$, a wingspan of $3.2m$, and a wing surface of $0.87m^2$. It has the same actuators configuration shown in Fig.~\ref{fig:actuators-config}~. A corresponding dynamic model was implemented in the Matlab\textregistered/Simulink\textregistered~environment, with the aerodynamics forces and torques modeled according to aerodynamic coefficients, and stability and control derivatives themselves estimated using the FlightStream\footnote[1]{Information available online at \href{https://www.darcorp.com/flightstream-aerodynamics-software/}{https://www.darcorp.com/flightstream-aerodynamics-software/} [retreived 28 August 2023]} aerodynamic modeling software.
The aerodynamic model used in these simulations is more elaborate than the analytical control model (in Eq.~(\ref{eq:fa-2})) that was used for control design. It is based on look-up tables that cover large angle of attack and side-slip domains and is complemented with flat-plate models outside the domain of convergence of the modeling software via sigmoid blending functions (similarly to Ref.~\cite{beard2012small} chapter 4, or Ref.~\cite{dronesaeroblending}). The thrust forces and shaft torques of the different blades are computed on the basis of a Blade-Element-Momentum-Theory (BEMT) model (see Ref.~\cite{bangura} or \cite{aircraft-control}, for example), which takes into account the effects of nonaxial airflow. Typical time constants and power limitations are also included in the dynamic modeling of the control surfaces and of the electrical motors.

\subsection{Practical Considerations}
\subsubsection{Estimation of the Air-Velocity Vector}

The peculiarity of UAVs is that they are usually not equipped with angle-of-attack and side-slip sensors. There is typically a pitot tube that measures a pressure difference translated into the airspeed component $v_{a,1}$. The control laws rely on the air-velocity vector $\bm v_a$, which requires estimating this vector despite the absence of angle of attack and sideslip sensors.
There are several methods to solve this problem; they are mainly classified into two categories, model-based estimators (see Ref.~\cite{wind-estimation2}, for example, or the method proposed in Ref.~\cite{2019-unified-automatica}) and kinematic estimators (see  Ref.~\cite{wind-estimation1}, for example). We propose here a simple kinematic method that exploits the measurement of the inertial speed $\bm v$ and assumes that the wind is horizontal and has no vertical component. It is also considered that side-slip remains low thanks to the passive stability of the airplane's geometry (weather vane effects via its vertical stabilizer and wing dihedral). The estimation error due to this assumption should be compensated by the robustness of the control terms thanks to the use of integral terms.
\newline
We denote the estimated air-velocity and inertial speed vectors by $\hat{\bm v}_a$, and $\hat{\bm v}$, respectively. The assumption of a horizontal wind implies that the vertical component of the air-velocity equals that of the inertial speed as follows:

\begin{equation}
\bm \hat{v}_a \cdot \bm k_0 = \hat{\bm v} \cdot \bm k_0
\end{equation}
Now, replacing $\bm v_a$ with its decomposition in the body fixed-frame gives:
\begin{equation}\label{proof-va-estimation}
\hat{v}_{a,1} ~\bm \imath \cdot \bm k_0 + \hat{v}_{a,2} ~\bm \jmath \cdot \bm \jmath_0 + \hat{v}_{a,3} ~\bm k \cdot \bm k_0  = \hat{\bm v} \cdot \bm k_0
\end{equation}

While the component $\hat{v}_{a,1}$ is considered to be obtained from the pitot tube measurement, we set $\hat{v}_{a,2}=0$ in accordance with the previous assumptions. The resulting expression for $\hat{v}_{a,3}$ is deduced from Eq.~\eqref{proof-va-estimation} as follows:

\begin{equation}
\hat{v}_{a,3}  = \frac{  \hat{\bm v} \cdot \bm k_0 - \hat{v}_{a,1} ~\bm \imath \cdot \bm k_0  } {\bm k \cdot \bm k_0  }
\end{equation}

The resulting approximation of $\bm v_a$ is  then $\hat{\bm v}_a=\hat{v}_{a,1} ~\bm \imath + \hat{v}_{a,3} ~\bm k$. It is of course necessary to avoid dividing by zero when $\bm k \cdot \bm k_0=0$; this can be done in practice by multiplying by $\frac{\bm k \cdot \bm k_0}{(\bm k \cdot \bm k_0)^2 + \epsilon}$ where $\epsilon$ is a small number, instead of the division. Moreover, this singularity occurs at bank angles near $90^{\circ}$. This flight condition is clearly outside the flight envelope of the vehicle especially during the transition maneuver and should be avoided by choosing proper limitations on desired acceleration terms in the \textit{speed vector control} module.

\subsubsection{Translation to Pulse-Width-Modulation (PWM) Signals}

The outputs of the flight controller for commercial UAVs are generally PWM electronic square signals, that are encoded in the autopilot software as values comprised between $1000 \mu s$ and $2000 \mu s$ representing the period of time of the \textit{on} state. The control terms that are computed need to be translated into corresponding PWM values. This means that for rotors the desired thrust value $t_i$ (or $T_{FW}$ for the pusher rotor) should be converted to the right PWM encoded value. The same applies for the deflection of control surfaces who are first computed as angles. In this research work, we first estimated the functions giving a control term $u$ as function of the applied PWM signal, in other words, $u=f_u (PWM_u)$; this was done by combining datasheet information with ground tests of the actuators. The next step consisted in inversing the previous relation, in other words, estimating the functions in the form $PWM_u=f_u^{-1} (u)$ that allows to deduce a command value $PWM_u$ for every desired actuation command $u$. Note that $f_u^{-1}$ can be obtained through a polynomial correlation, or even a tabulated interpolation.

\subsection{Control parameters and simulation results}

The control parameters used in the control laws are shown in Table \ref{tab:params}~. One can notice that a single set of parameters is applied for the entire flight envelope without the need for gain scheduling. This is made possible by the nonlinear aspect of the control design which is model based and takes into account instantaneous values of speed, airspeed and dynamic pressure (in Eqs.~\eqref{eq:def-alat-pf-cont}, \eqref{eq:th-d}, \eqref{eq:th-e} and \eqref{eq:alloc-fw-inv}, for instance). However, this is not guaranteed to be the case for all applications, especially for larger vehicles with lower actuator bandwidth. Typically, stability, performance and robustness analysis are peformed to validate parameters tuning and scheduling. Such assessment methods may involve frequency-domain analysis such as evaluation of stability margins  and disturbance rejection bandwidth, or time-domain analysis using Monte-Carlo simulations to assess robustness in the presence of parameter dispersions. Some design metrics attained for the presented vehicle and control parameters are briefly presented in Appendix \ref{sec:app3}. We refer the interested reader to the abundant literature on the subject (see, for example, Refs.~\cite{tischler2017practical,handling_qualities_VTOL_Berger,HQMC}), and we focus in this paper on presenting the contributions of the proposed control approach.

 The parameter $ \lambda \in [0,1]$, which represents the blending coefficient for torque actuation, was varied smoothly and linearly. During phase T2, it was increased from $0$ to $1$ in 2 seconds according to the expression $\lambda(\Delta t)= min(0.5 \Delta t, 1)$, with $\Delta t$ the elapsed time since the beginning of T2. During phase BT3, it was decreased from $1$ to $0$ in 1 second according to the expression $\lambda(\Delta t)= max(1-\Delta t, 0)$, with $\Delta t$ the elapsed time since the beginning of BT3.
\begin{table}[hbt!]
\caption{\label{tab:params} Control parameters}
\begin{tabular}{ll}
\hline
\textbf{Control Module}& \textbf{Gains} \\\hline
Altitude Control& $k_z=0.25$, $v_{z,max}=1$, $v_{z,min}=-1.5$ \\ \hline
Guidance, trajectory tracking& $k_p=0.29$, $v_{h,max}=5$   \\ \hline
Vertical speed control& $k_{v,z}=3.65$, $k_{I,vz}=1.25$, $a_{z,max}=4.5$, $a_{z,min}=-5.5$, $\Delta_{I,vz}=3.15$   \\ \hline
Horizontal speed control& $k_{v,h}=1.5$, $k_{I,vh}=0.7$, $a_{h,max}=3.35$, $\Delta_{I,vh}=2.75$   \\ 
Velocity tracking & \\ \hline
Horizontal speed control& $k_{t}=2.4$, $k_{I,t}=1.1$, $a_{t,max}=5$, $a_{t,min}=-1$, $\Delta_{I,vt}=1.3$, $k_h=0.8$, $k_{I,h}=0.16$, \\
Path following   & $a_{l,max}=5.21$, $\Delta_{I,h}=1.5$  \\ \hline
Model parameters& $m=17.5$, $\rho=1.2$, $c_0=0.074$, $\bar{c}_0=5.074$, $S=0.868$, $\alpha_0=4.53^\circ=0.0791~ rad$  \\ \hline
Attitude control& $k_{\bm \imath}=6$, $k_{\bm \jmath}=6$, $k_{\bm k}=1.8$  \\ \hline
Angular rates control& $k_{p,\omega_1}=11$, $k_{p,\omega_2}=12$, $k_{p,\omega_3}=4.75$, $k_{I,\omega_1}=10$, $k_{I,\omega_2}=25$, $k_{I,\omega_3}=0.15$, \\
                           & $\Delta_{I,\omega_1}=3.5$, $\Delta_{I,\omega_2}=8$, $\Delta_{I,\omega_3}=0.5$ \\ \hline
Actuators and inertia& $d=0.55$, $e=0.55$, $f=0.025$, $b=3.2$, $c=0.3$, $\eta=0.021$, $C_{l,\delta_a}=0.002$,\\
&  $C_{l,\delta_{rel}}=C_{l,\delta_{rer}}=0$, $C_{m,\delta_a}=0$, $C_{m,\delta_{rel}}=C_{m,\delta_{rer}}=0.006$, \\
  & $C_{n,\delta_{rer}}=-C_{n,\delta_{rel}}=0.0018$, $J=diag(0.87,1.11,1.84)$ \\ \hline
Transition parameters & $\theta_{T0}=0^\circ$, $v_{z,T0}=-1 m/s$, $\bm v_{hor,r}= 5 m/s$, $\theta_{T1}=0^\circ$, $v_{z,T1}=-1.1 m/s$,\\
                      & $v_{a,T1}=9 m/s$, $\theta_{T2}=0^\circ$, $v_{z,T2}=-0.9 m/s$, $\theta_{T3}=3^\circ$, $v_{z,T3}=0 m/s$,\\ 
                      & $v_{a,FW}= 20 m/s$
 \\ \hline
Back-Transition parameters & $v_{z,BT0}=0.5 m/s$, $v_{a,FW}= 20 m/s$, $\theta_{BT1}=3^\circ$, $v_{z,BT1}=0 m/s$\\
   & $v_{z,BT2}=0.12 m/s$, $v_{a,BT2}= 10 m/s$, $\theta_{BT3}=3^\circ$\\
   &$z_{BT3}$, {\small{and}} $z_{BT4}$ {\small{are set to the initial altitudes at the beginning of BT3 and BT4 respectively.}} \\ \hline
\hline

\end{tabular}
\end{table}

The chosen flight scenario consists of five consecutive parts:
\begin{enumerate}
\item MultiCopter (MC) mode: this phase consists of a manual takeoff until hovering at a fixed altitude, with the pilot sending position and inertial speed commands and the control modules running in a multicopter mode, neglecting aerodynamic compensation, and imposing a thrust direction, in other words, $\gamma_{T,r}=-\frac{\pi}{2}$, and a torque control achieved throught differential thrust with $\lambda=0$~.
\item Transition:  here the pilot triggers the transition (phases T0 to T4) which is achieved autonomously along the initial heading.
\item Fixed-Wing (FW) mode: in this phase the pilot is commanding the vehicle by sending airspeed, heading and altitude (or climb and descent rate) setpoints. The control modules are running in a fixed-wing mode, taking into account aerodynamic compensation, and imposing a thrust direction, in other words, $\gamma_{T,r}=0$, and a torque control achieved through control surfaces with $\lambda=1$~.
\item Back-Transition: here the pilot triggers the back-transition (phases BT0 to BT4), which is achieved autonomously along the intial heading.
\item Multi-Copter (MC) mode: in this phase the pilot is again taking control similarly to the first phase, with the intention to manually land the vehicle, with the control modules imposing $\gamma_{T,r}=-\frac{\pi}{2}$ and $\lambda=0$~.
\end{enumerate}

A steady wind is added to the simulation, in such a manner that during the transition maneuver a front wind is blowing at $3m/s$, which becomes a rear wind during the back transition. Both the transition and back-transition experience a lateral wind blowing at $1m/s$. To further test the robustness of the control, the simulated weight of the vehicle has been increased to $19 Kg$ while the value \textit{viewed} by the control modules remained at $m=17.5 Kg$.

\begin{figure}[hbt!]
\centering
\includegraphics[width=0.75\textwidth]{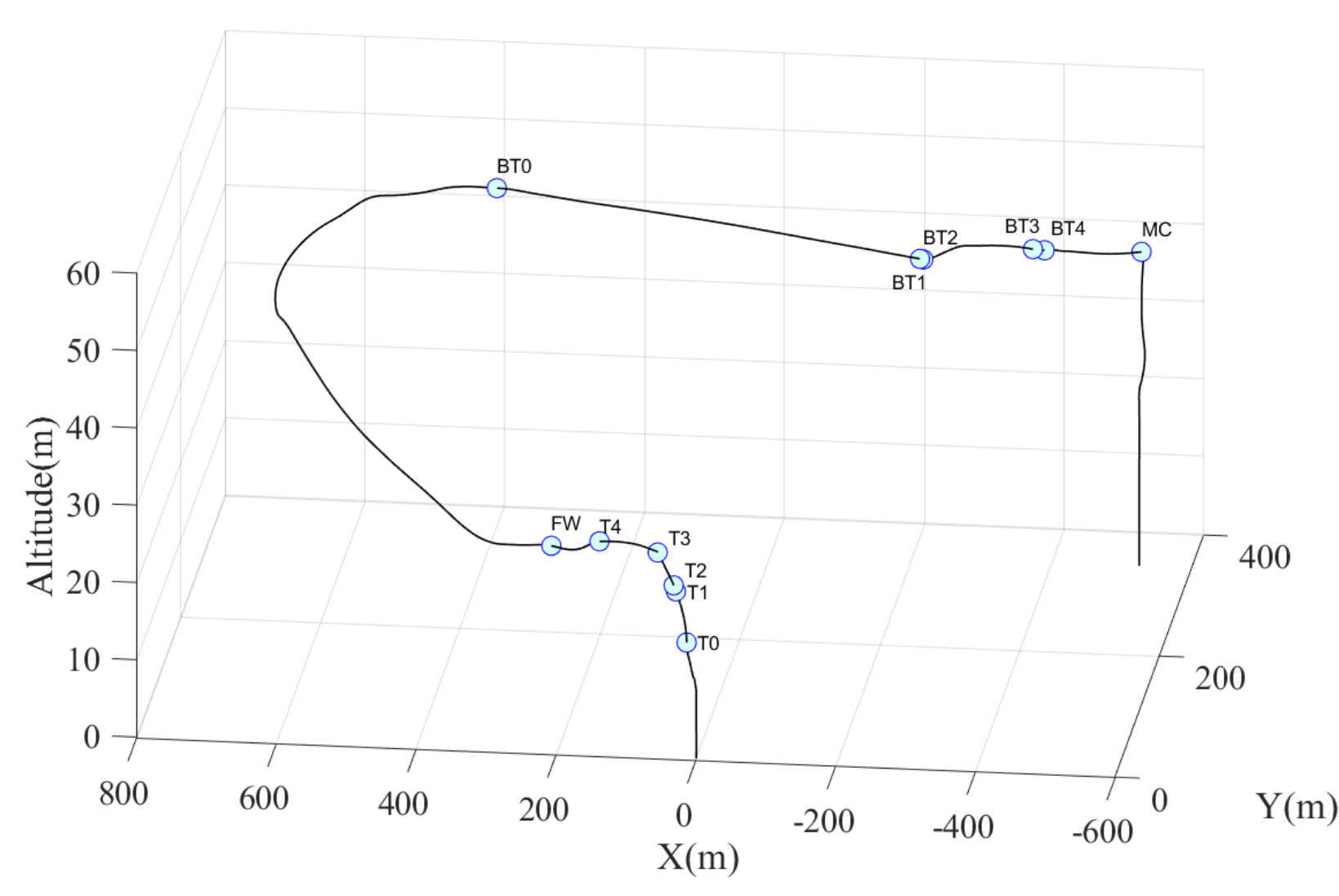}
\caption{Aircraft's trajectory.}
\label{fig:HIL-3D}
\end{figure}

\begin{figure}[H]
\centering
\begin{minipage}{.49\textwidth}
\centering
\includegraphics[width=1.0\textwidth]{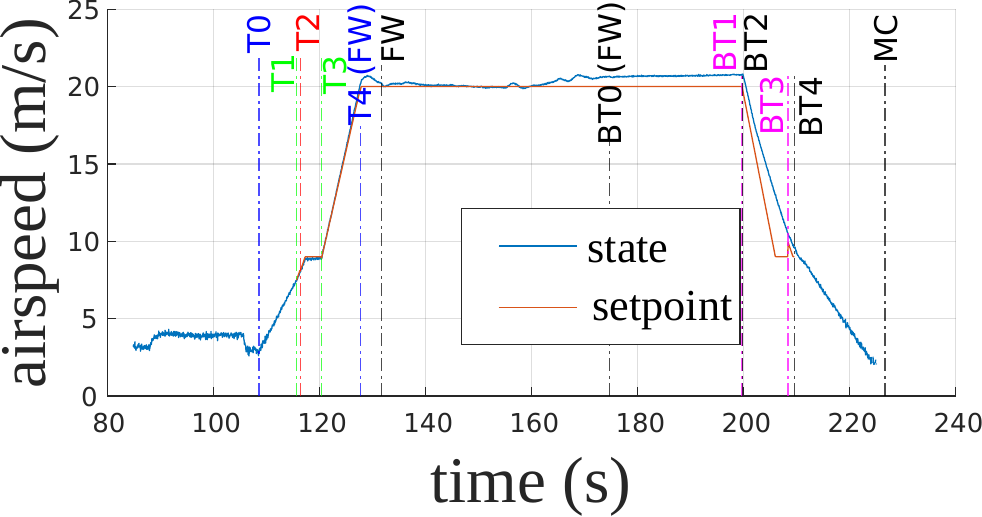}
\caption{Airspeed.}
\label{fig:HIL-airspeed}
\end{minipage}
\begin{minipage}{.49\textwidth}
\centering
\includegraphics[width=1.0\textwidth]{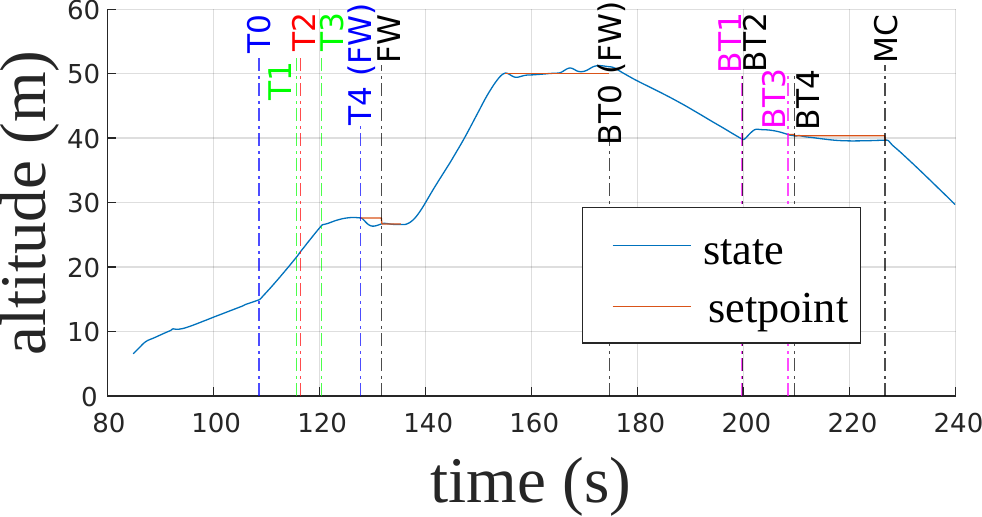}
\caption{Altitude.}
\label{fig:HIL-altitude}
\end{minipage}
\end{figure}

\begin{figure}[hbt!]
\centering
\begin{minipage}{.49\textwidth}
\centering
\includegraphics[width=1.0\textwidth]{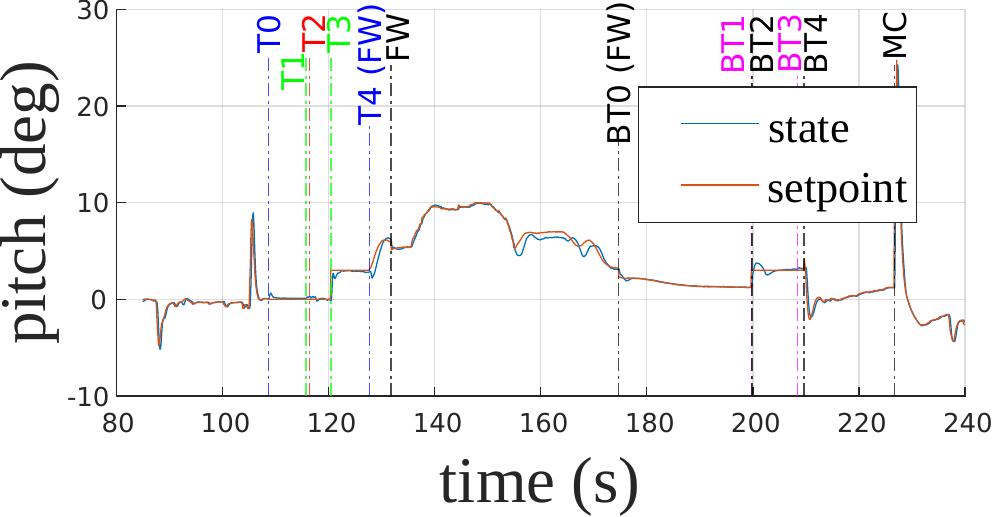}
\caption{Pitch angle.}
\label{fig:HIL-pitch}
\end{minipage}
\begin{minipage}{.49\textwidth}
\centering
\includegraphics[width=1.0\textwidth]{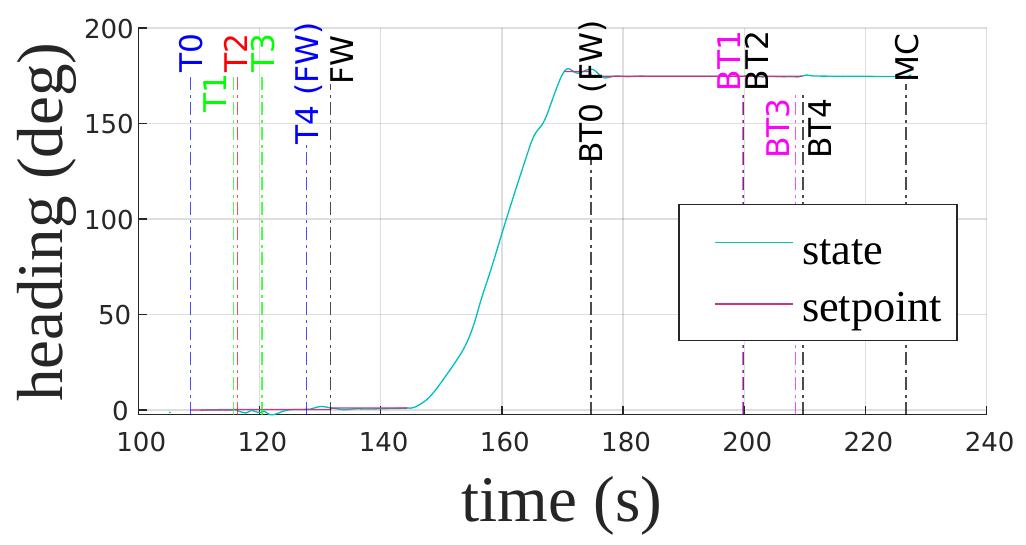}
\caption{Heading of the vehicle with respect to  axis $\bm \imath_0$}
\label{fig:HIL-heading}
\end{minipage}
\end{figure}

\begin{figure}[hbt!]
\centering
\includegraphics[width=0.49\textwidth]{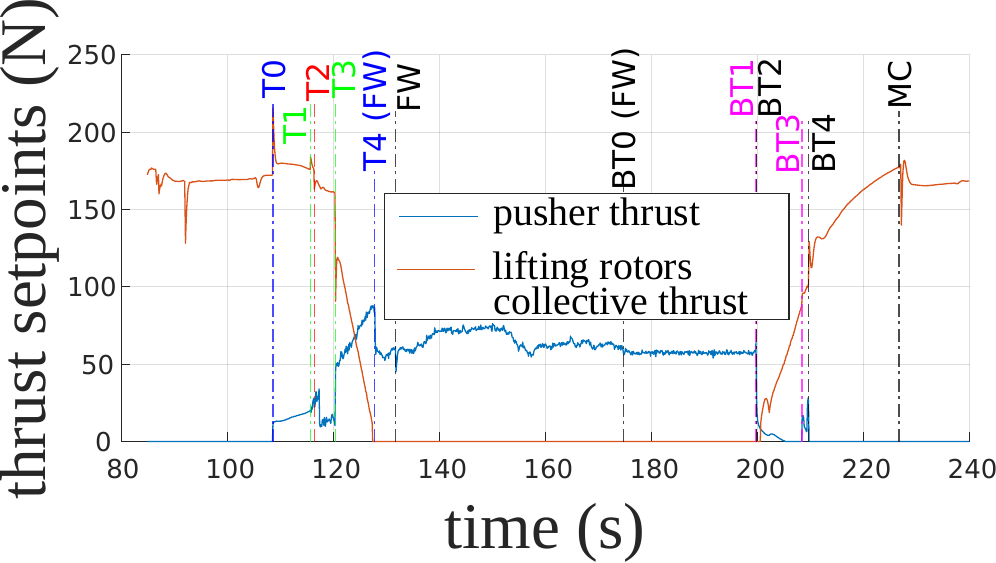}
\caption{Collective multicopter thrust ($|\bm T_{MC,r}|$) and pusher thrust ($|\bm T_{FW,r}|$) commands.}
\label{fig:HIL-cont-thrusts}
\end{figure}

\begin{figure}[hbt!]
\centering
\begin{minipage}{.49\textwidth}
\centering
\includegraphics[width=1.0\textwidth]{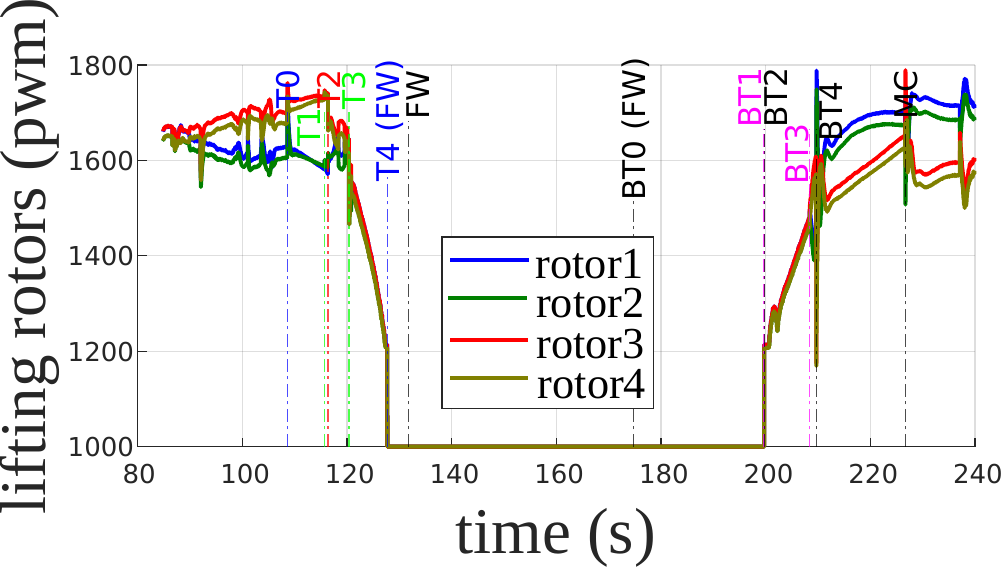}
\caption{Lifting rotors PWM commands.}
\label{fig:HIL-cont-lift-rotors}
\end{minipage}
\begin{minipage}{.49\textwidth}
\centering
\includegraphics[width=1.0\textwidth]{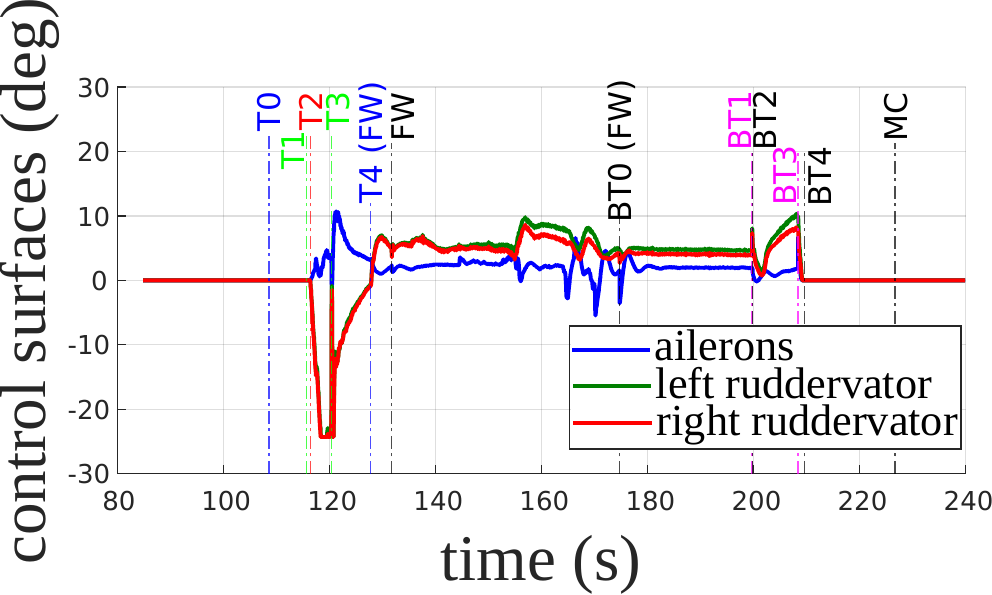}
\caption{Control surfaces commands.}
\label{fig:HIL-cont-surfaces-deg}
\end{minipage}
\end{figure}

Simulation results reported in Figs.~\ref{fig:HIL-3D}-\ref{fig:HIL-cont-surfaces-deg}, show that the control modules successfully achieved the transition and back-transition maneuvers progressively. The different setpoints were tracked with acceptable errors magnitudes in all the phases, as well as in manually piloted MC and FW phases. We notice that the desired deceleration of the airspeed during phase BT2 cannot be achieved at the required rate, that is due to the physical impossibility to achieve negative thrust with the pusher rotor; however, this does not preclude the pursuit of the back-transition and the deceleration of the vehicle toward a hover flight condition.

\section{Flight Experiment}\label{sec:test}

\begin{figure}[H]
\centering
\includegraphics[width=0.35\textwidth]{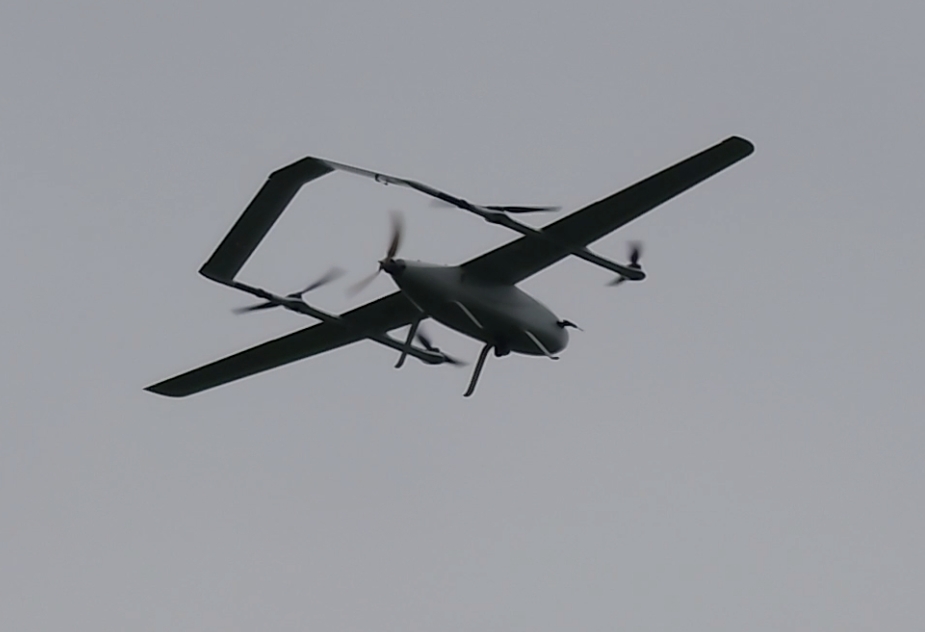}
\caption{The VTOL UAV during the experiment.}
\label{fig:test-vehicle}
\end{figure}

In this section, we experimentally test the transition maneuvers, with the flight controller integrated on board the compound VTOL aircraft shown in Fig.~\ref{fig:test-vehicle}. During the experiment, a wind with an average magnitude of $3 m/s$ was blowing from the east.  After performing a manual takeoff in a multicopter configuration, the pilot aligned the vehicle to face the coming wind and triggered the transition phase. Then, after having reached the final step (T4) of the transition in which the vehicle was flying autonomously in a fixed-wing configuration, the pilot initiated the back-transition maneuver. The vehicle was eventually decelerated to a hovering flight, after which it was manually landed.

\begin{figure}[H]
\centering
\includegraphics[width=0.7\textwidth]{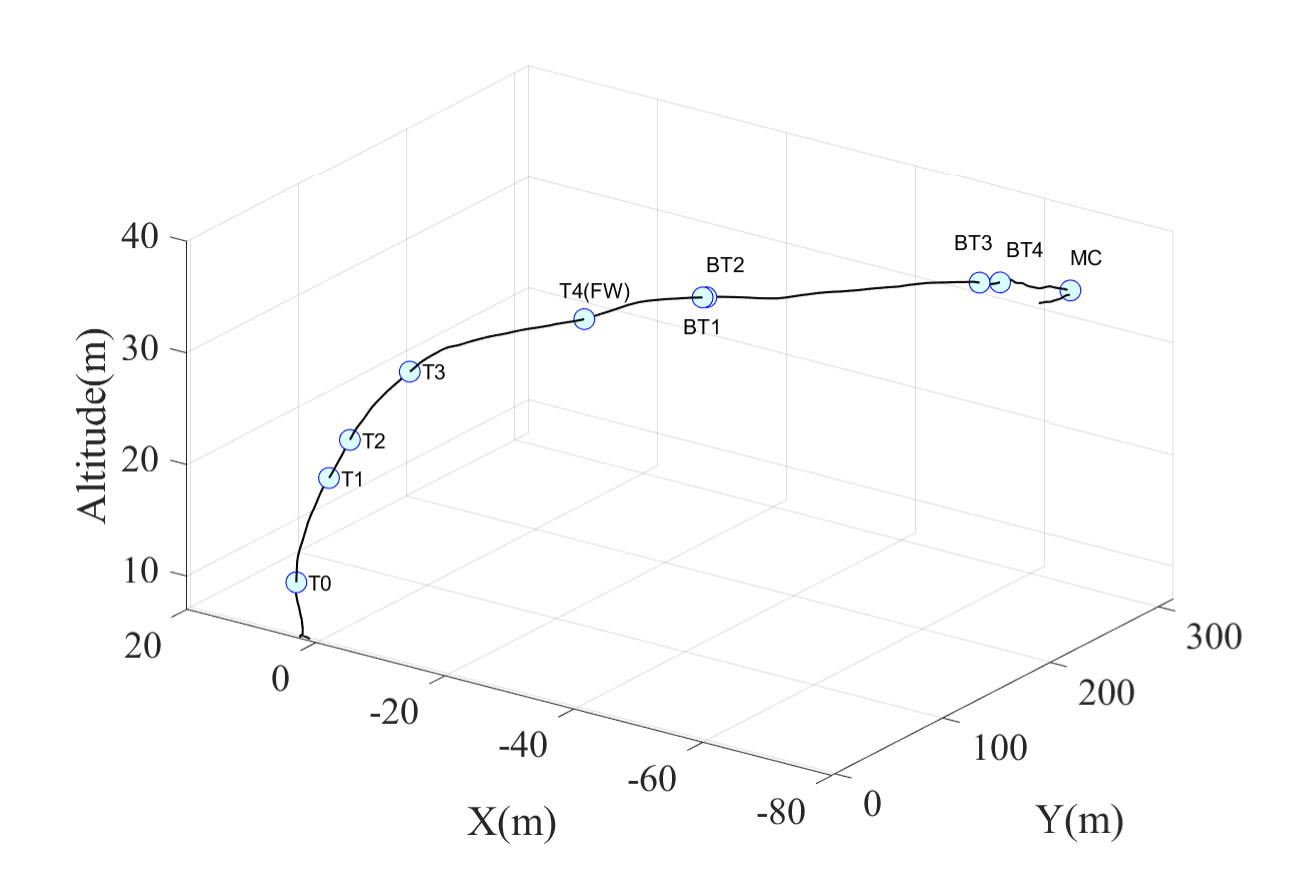}
\caption{Aircraft's trajectory.}
\label{fig:test-3D}
\end{figure}

\begin{figure}[H]
\centering
\begin{minipage}{.49\textwidth}
\centering
\includegraphics[width=1.0\textwidth]{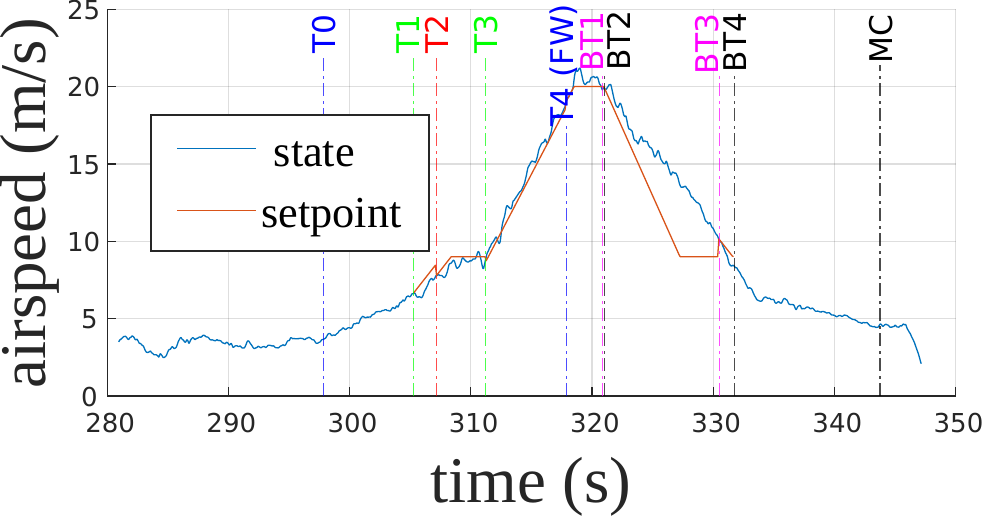}
\caption{Airspeed.}
\label{fig:test-airspeed}
\end{minipage}
\begin{minipage}{.49\textwidth}
\centering
\includegraphics[width=1.0\textwidth]{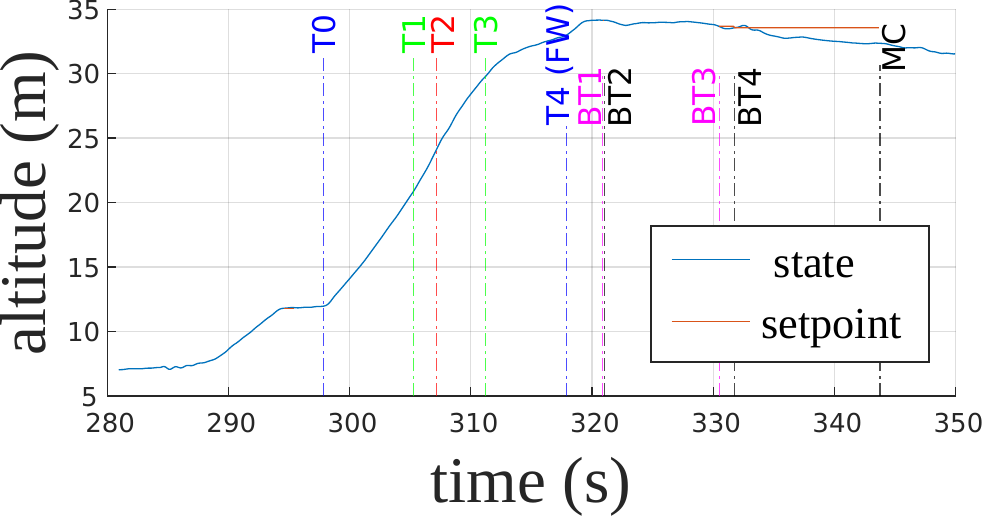}
\caption{Altitude.}
\label{fig:test-altitude}
\end{minipage}
\end{figure}

\begin{figure}[H]
\centering
\begin{minipage}{.49\textwidth}
\centering
\includegraphics[width=1.0\textwidth]{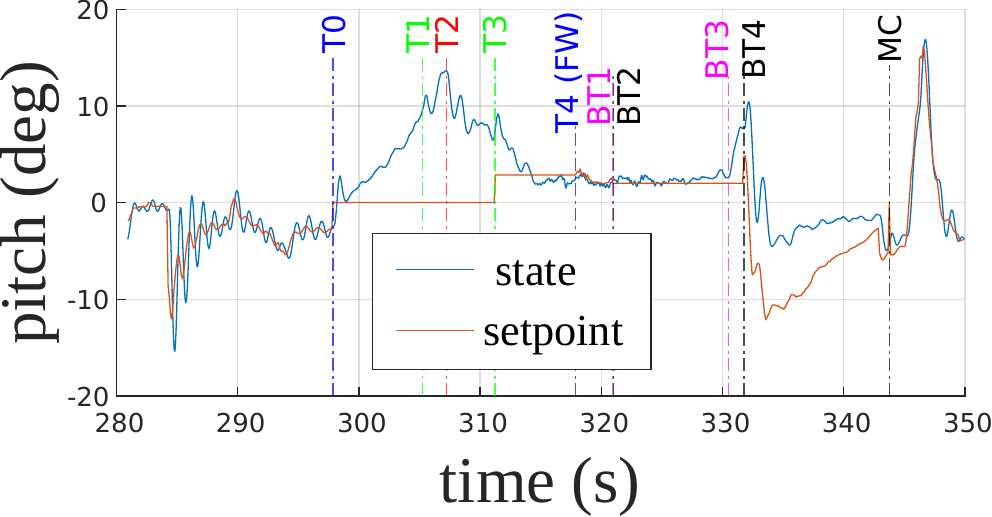}
\caption{Pitch angle.}
\label{fig:test-pitch}
\end{minipage}
\begin{minipage}{.49\textwidth}
\centering
\includegraphics[width=1.0\textwidth]{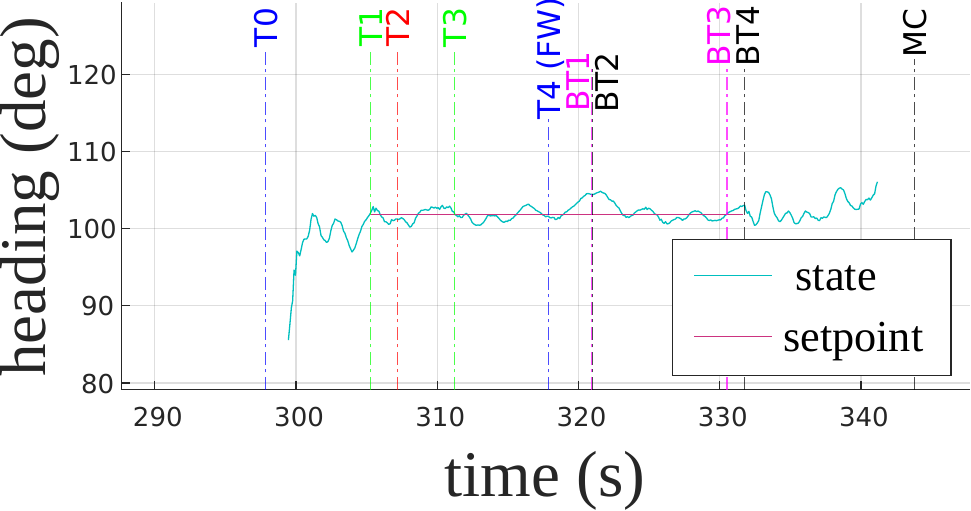}
\caption{Heading of the vehicle with respect to  axis $\bm \imath_0$}
\label{fig:test-heading}
\end{minipage}
\end{figure}

\begin{figure}[H]
\centering
\includegraphics[width=0.49\textwidth]{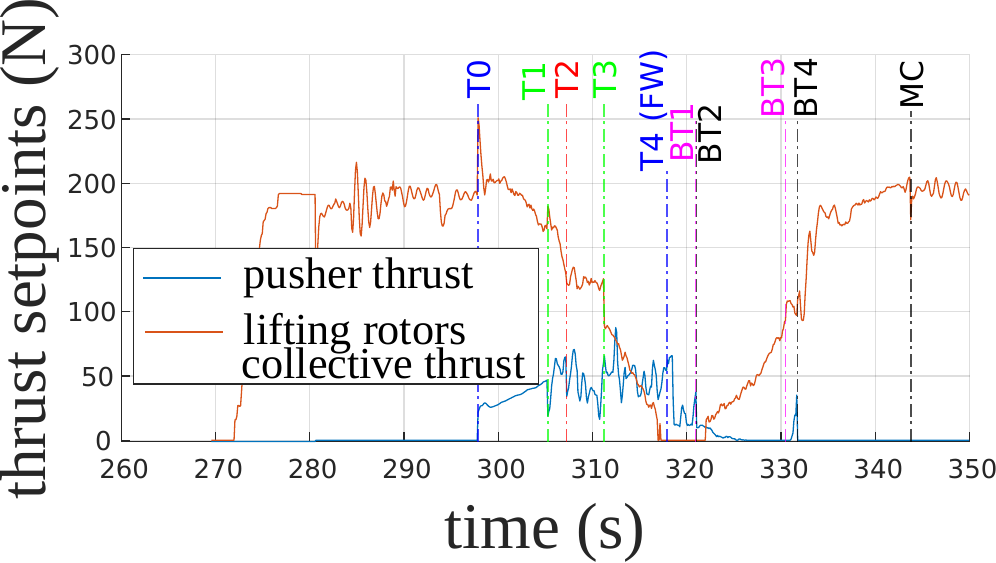}
\caption{Collective Multi-Copter thrust ($|\bm T_{MC,r,}|$) and pusher thrust ($|\bm T_{FW,r,}|$) commands.}
\label{fig:test-cont-thrusts}
\end{figure}

\begin{figure}[H]
\centering
\begin{minipage}{.49\textwidth}
\centering
\includegraphics[width=1.0\textwidth]{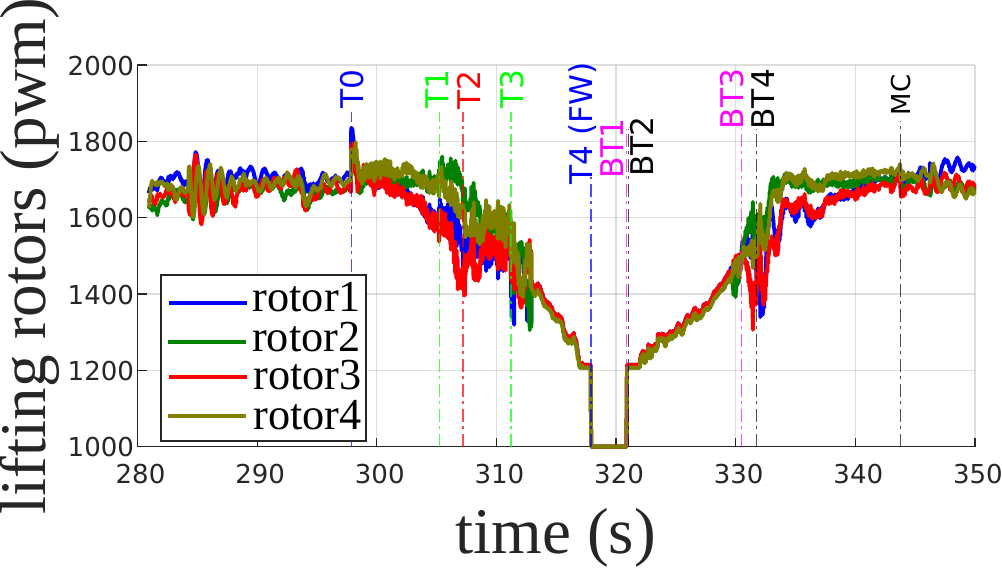}
\caption{Lifting rotors PWM commands.}
\label{fig:test-cont-lift-rotors}
\end{minipage}
\begin{minipage}{.49\textwidth}
\centering
\includegraphics[width=1.0\textwidth]{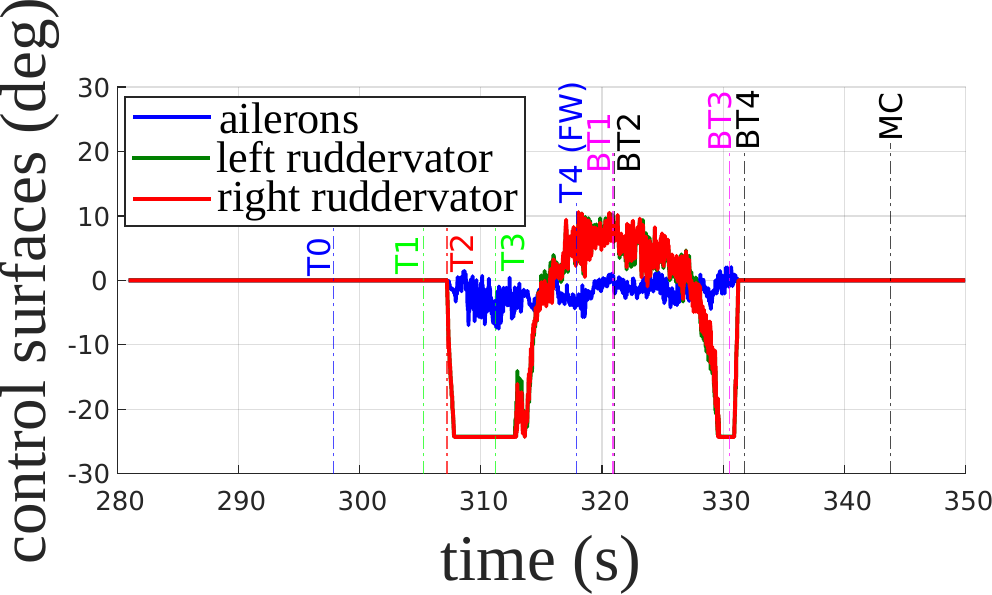}
\caption{Control surfaces commands.}
\label{fig:test-cont-surfaces-deg}
\end{minipage}
\end{figure}

Figures ~\ref{fig:test-3D}-\ref{fig:test-cont-surfaces-deg} show that the transition strategy and control laws succeeded in achieving the reconfiguration of the vehicle between the multi-copter and fixed-wing configurations. The desired heading is tracked with an error of less than $3^\circ$, which has resulted in an almost longitudinal flight respecting a narrow flight corridor. Also, note that no loss of altitude is experienced during the transition, and the airspeed is accelerated in accordance with the varying airspeed setpoint. A tracking error on the desired pitch can be noticed during the beginning of the transition and the end of the back-transition\footnote[1]{This could disrupt the rate of climb setpoint tracking and will even make it more complicated to achieve a fixed altitude transition.}. This was attributed to downwash effects on the rear lifting rotors when flying at high airspeeds, which decreases their efficiency and leads to a positive torque offset\footnote[2]{This effect was not observed during the HIL simulations. Indeed the simulation of the previous section did not take into account such downwash effects due to the complexity of their modeling. We aim in this flight experiment to evaluate the robustness of the laws even in the face of unmodeled effects that are expected when working on such VTOL configurations.}. This highlights the complexity of designing such VTOL configurations that has to take into account complex aerodynamic interactions. However, the pitch tracking error did not prevent the vehicle from achieving the transition by accelerating to higher airspeeds at which aerodynamic control surfaces become sufficiently efficient to restore control robustness. 

\section{Conclusion}\label{sec:conclusion}

A flight control design and a transition strategy for compound VTOL aircraft was presented. The control laws are generic and applicable to any vehicle configuration, whether it is a multicopter, a fixed-wing or a VTOL aircraft with additional actuation capabilities and multiple means for torque generation. This control design was detailed for the compound VTOL configuration and was associated with a sequence of high-level setpoints in order to constitute a transition strategy that exploits the convergence and stability properties of the control laws throughout the intermediate transition states. Simulation results and flight experiments show the strategy's potential to provide stable transition maneuvers, which makes it eligible to satisfy narrow flight corridor requirements and adds the capability to abandon the transition and back transition phases at any intermediate point of the maneuvers.

\section*{Appendix: Convergence and Stability Analysis}

\subsection{Proof for the thrust vector and attitude setpoints computation}\label{sec:app1}

Here, we give proofs for the expressions corresponding to Eqs.~\eqref{cont-imp-thrust-dir}, \eqref{cont-imp-pitch} and \eqref{cont-thr}. The following proof differs from the work in Ref.~\cite{2019convertible} by computing the orientation of the desired body axis with respect to the apparent acceleration $\bm a'$ instead of the air-velocity vector $\bm v_a$. This allows us to define the control terms even when $\bm v_a$ is equal to $\bm{0}$. Another difference with Ref.~\cite{2019convertible}, is the possibility to impose a pitch angle, as reported in the second case.
Recall that the objective is to solve the following equation:
\begin{equation}\label{eq:proof1}
\bm a'= \frac{{\bm F}_a}{m} + \frac{\bm T_r}{m}
\end{equation}
We report here the previous definitions of the aerodynamic force model $\bm F_a$  and the vectors $\bm a'$, $\bm d$ and $\bm e$:
\begin{align}
\bm F_a &= -\frac{1}{2} \rho S |\bm v_a| (c_0 (\bm v_a \cdot \bm \imath_{2r}) \bm \imath_{2r} + \barbar{c}_0 (\bm v_a \cdot \bm \jmath_r) \bm \jmath_r+\bar{c}_0 (\bm v_a \cdot \bm k_{2r}) \bm k_{2r}) \label{eq:proof2} \\
\bm a' &= \bm a_r - \bm g \label{eq:proof201} \\
\bm d &= m \bm a' + \frac{1}{2} \rho S |\bm v_a|c_0 \bm v_a \label{eq:proof202}\\
\bm e &= m \bm a' + \frac{1}{2} \rho S |\bm v_a| \bar{c}_0 \bm v_a \label{eq:proof203} 
\end{align}
Replacing these expressions in Eq.~\eqref{eq:proof1} yields
\begin{equation}\label{eq:proof3}
(\bm d \cdot\bm  \imath_{2r} - |\bm T_r| \cos(\gamma_{T,r} + \alpha_0) ) \bm \imath_{2r} + (\bm e \cdot \bm k_{2r} - |\bm T_r| \sin(\gamma_{T,r} + \alpha_0)) \bm k_{2r} + (\bm a' \cdot \bm \jmath_r + \frac{1}{2} \rho S |\bm v_a| \barbar{c}_0 v_{a,2}) \bm \jmath_r = 0
\end{equation}
With the chosen expressions of $\bm \jmath_r$ in Sec.~\ref{section-thrust-vectoring}, and especially the expression employed in the presence of aerodynamic forces, in other words, Eq.~\eqref{eq:jr2}, the last term in Eq.~\eqref{eq:proof3} equals $\bm{0}$.
The equality in Eq.~\eqref{eq:proof3} becomes equivalent to solving the following system:
\begin{equation}\label{eq:proof4}
\begin{cases}
\bm d \cdot\bm  \imath_{2r} - |\bm T_r| \cos(\gamma_{T,r} + \alpha_0)  & = 0\\
\bm e \cdot \bm k_{2r} - |\bm T_r| \sin(\gamma_{T,r} + \alpha_0) &= 0
\end{cases}
\end{equation}
Which is also equivalent to
\begin{equation}\label{eq:proof5}
\begin{cases}
(\bm d  \cdot \bm  \imath_{2r}) \sin(\gamma_{T,r} + \alpha_0)  - (\bm e \cdot \bm k_{2r}) \cos(\gamma_{T,r} + \alpha_0) = 0 \\
|\bm T_r| = \cos(\gamma_{T,r} + \alpha_0) (\bm d \cdot\bm  \imath_{2r}) + \sin(\gamma_{T,r} + \alpha_0) (\bm e \cdot \bm k_{2r})
\end{cases}
\end{equation}
With the additional degree of freedom provided by the thrust direction term $\gamma_T$, there exists an infinity of solutions for Eq.~\eqref{eq:proof5}. We restrict the analysis for the two cases considered in this paper. 
\begin{itemize}
\item \textbf{case 1:} Imposed thrust direction $\gamma_r$
We recall the definition of $\bm a^{' \perp}$:
\begin{equation}
\bm a^{'\perp} = \bm a'  \times \bm \jmath_r
\end{equation}
We consider that $\bm \imath_r$ can be obtained by rotating $\frac{\bm a'}{|\bm a'|}$ around $\bm \jmath_r$  with an angle $\gamma$ as shown in Fig.~\ref{fig:axis-proof}. 
\begin{figure}[hbt!]
\centering
\includegraphics[width=.5\textwidth]{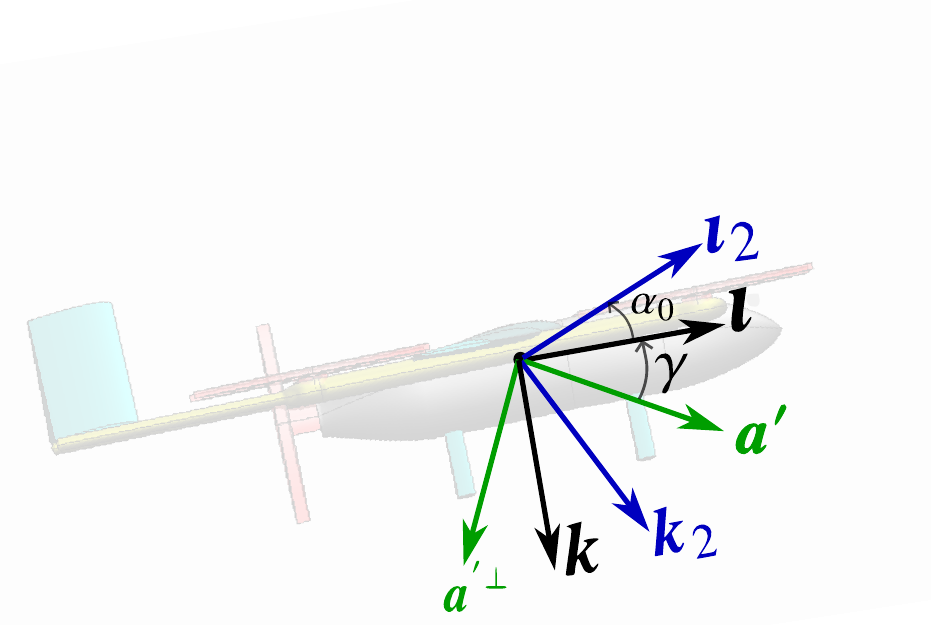}
\caption{Definitions of angles $\gamma$ and $\alpha_0$. Note that the shown angles are positive according to the chosen convention.}
\label{fig:axis-proof}
\end{figure}
This gives
\begin{equation}\label{eq:proof6}
\begin{cases}
\bm \imath_r &= \cos{\gamma} \frac{\bm a'}{|\bm a'|} - \sin{\gamma} \frac{\bm a^{'\perp}}{|\bm a^{'\perp}|} \\
\bm k_r &= \sin{\gamma} \frac{\bm a'}{|\bm a'|} + \cos{\gamma} \frac{\bm a^{'\perp}}{|\bm a^{'\perp}|} 
\end{cases}
\end{equation}
The solution to the problem can then be given by
\begin{equation}\label{eq:proof7}
\begin{cases}
\bm k_r &= \sin{\gamma} \frac{\bm a'}{|\bm a'|} + \cos{\gamma} \frac{\bm a^{'\perp}}{|\bm a^{'\perp}|} \\
\bm \imath_r &=\bm \jmath_r \times \bm k_r
\end{cases}
\end{equation}
From here, it suffices to solve the first equation in Eq.~\eqref{eq:proof5} for $\gamma$.
Given the previous definition of angle $\alpha_0$ in Sec.~\ref{sec-aero}, we can write the following:
\begin{equation}\label{eq:proof8}
\begin{cases}
\bm \imath_{2r} &= \cos({\gamma+\alpha_0}) \frac{\bm a'}{|\bm a'|} - \sin({\gamma+\alpha_0}) \frac{\bm a^{'\perp}}{|\bm a^{'\perp}|} \\
\bm k_{2r} &= \sin({\gamma+\alpha_0}) \frac{\bm a'}{|\bm a'|} + \cos({\gamma+\alpha_0}) \frac{\bm a^{'\perp}}{|\bm a^{'\perp}|} 
\end{cases}
\end{equation}
Replacing Eq.~\eqref{eq:proof8} in the first Eq. of ~\eqref{eq:proof5}, we get:
\begin{equation}\label{eq:proof9}
\begin{cases}
\cos(\gamma+\alpha_0) ( \sin(\gamma_{T,r}+ \alpha_0)   ( \bm d \cdot \bm a') - \cos(\gamma_{T,r}+\alpha_0) \bm e \cdot \bm a^{'\perp} )  \\
- \sin(\gamma + \alpha_0) ( \sin(\gamma_{T,r} + \alpha_0) \bm d \cdot \bm a^{'\perp} + \cos(\gamma_{T,r} + \alpha_0) \bm e \cdot  \bm a') = 0
\end{cases}
\end{equation}
or equivalently
\begin{equation}\label{eq:proof10}
\begin{cases}
\gamma &= atan2(y,x) - \alpha_0 \\
\text{with,}\\
y &= \sin(\gamma_{T,r} + \alpha_0) \bm d \cdot \bm a' - \cos(\gamma_{T,r} + \alpha_0) \bm e \cdot \bm a^{'\perp} \\
x&= \cos(\gamma_{T,r} + \alpha_0) \bm e \cdot \bm a' + \sin(\gamma_{T,r} + \alpha_0) \bm d \cdot \bm a^{'\perp}
\end{cases}
\end{equation}
~\\
\item \textbf{case 2:} Imposed pitch angle $\theta_r$:

First, we recall the definition of the horizontal unitary vector $\bm \eta$ that is orthogonal to $\bm \jmath_r$:
\begin{equation}
\bm \eta =\frac{\bm \jmath_r \times \bm k_0}{|\bm \jmath_r \times \bm k_0|} 
\end{equation}
The unitary vector $\bm \eta^{\perp}$ that is orthogonal to both $\bm \eta$ and $\bm \jmath_r$ is given by
\begin{equation}
\bm \eta^{\perp} =\frac{\bm \jmath_r \times \bm \eta}{|\bm \jmath_r \times \bm \eta|}
\end{equation}
Because $\bm \eta$ is horizontal, an expression for the body axis $\bm \imath_r$ that corresponds to a pitch angle of $\theta_r$ can be expressed in the frame $(\bm \eta,\bm \eta^{\perp})$ as
\begin{equation}
\bm \imath_r = \cos(\theta_r) \bm \eta + \sin(\theta_r) \bm \eta^{\perp}
\end{equation}
Then, $\bm k_r$ can be deduced as $\bm k_r=\bm \imath_r \times \bm \jmath_r$.
It remains to find the solution for the thrust angle $\gamma_{T,r}$. The first equation in \eqref{eq:proof5}, is equivalent to
\begin{equation}\label{eq:proof11}
\gamma_{T,r} + \alpha_0 = atan2(\bm e \cdot \bm k_{2r}, \bm d \cdot \bm \imath_{2r})
\end{equation}
Using the expressions of the modified body axis in Eq.~\eqref{eq:modified-body-axis}, and replacing them in Eq.~\eqref{eq:proof11}, we find the expression for the desired thrust direction $\gamma_{T,r}$ as expressed in Eq.~\eqref{cont-imp-pitch}.
\end{itemize}
For the two preceding cases (imposed thrust direction and imposed pitch angle), the solution for the thrust norm $|\bm T_r|$ as function of the desired  body axis $(\bm \imath_r, \bm k_r)$ can be obtained by combining Eq.~\eqref{eq:modified-body-axis} with the second equation in \eqref{eq:proof5} to give the expression in Eq.~\eqref{cont-thr}.

\subsection{Convergence analysis for the translation dynamics}\label{sec:app12}

Here, we analyse the convergence of the translation dynamics. The objective is to reduce the closed-loop equation of the translation dynamics to the form

\begin{equation}
\dot{\bm v}=\bm a_r + \bm o(t), ~~ \displaystyle{\lim_{t \to \infty}} \bm o(t)=\bm{0}
\end{equation} 
with $\bm a_r$ a feedback control law equivalent to a desired acceleration vector that asymptotically (and locally exponentially) stabilizes the position and velocity tracking errors at $\bm{0}$, and $\bm o(t)$ a sum of additive perturbation terms that asymptotically (and locally exponentially) converges to $\bm{0}$.

From Eq.~(\ref{eq:trans-dyn}) we have
\begin{equation}\label{eq:proof501}
	\dot{\bm v} ={\bm g} + \frac{{\bm F}_a}{m} + \frac{\bm T}{m}
\end{equation}
By combining the definition of $\bm a'$ from Eq.~(\ref{eq:proof201})  with Eq.~(\ref{eq:proof501}), one can write:
\begin{equation}\label{eq:proof502}
		\dot{\bm v} =\bm a_r - \bm a' + \frac{{\bm F}_a}{m} + \frac{\bm T}{m}
\end{equation}
We report again the definitions of the aerodynamic force model $\bm F_a$, the thrust vector $\bm T$ and the vectors $\bm d$ and $\bm e$:
\begin{align}
\bm F_a &= -\frac{1}{2} \rho S |\bm v_a| (c_0 (\bm v_a \cdot \bm \imath_{2}) \bm \imath_{2} + \barbar{c}_0 (\bm v_a \cdot \bm \jmath) \bm \jmath+\bar{c}_0 (\bm v_a \cdot \bm k_{2}) \bm k_{2}) \label{eq:proof503} \\
\bm T&= |\bm T| (\cos(\gamma_T + \alpha_0) \bm \imath_2 + \sin(\gamma_T+ \alpha_0) \bm k_2) \label{eq:proof504} \\
\bm d &= m \bm a' + \frac{1}{2} \rho S |\bm v_a|c_0 \bm v_a \label{eq:proof505}\\
\bm e &= m \bm a' + \frac{1}{2} \rho S |\bm v_a| \bar{c}_0 \bm v_a \label{eq:proof506} 
\end{align}
Replacing these expressions in Eq.~\eqref{eq:proof502} yields:
\begin{equation}\label{eq:proof507}
	\dot{\bm v} =\bm a_r -\frac{1}{m}(\bm d \cdot \bm \imath_2 -|\bm T| \cos(\gamma_T + \alpha_0) ) \bm \imath_2 -\frac{1}{m}(\bm e \cdot \bm k_2 -|\bm T| \sin(\gamma_T + \alpha_0) ) \bm k_2 -\frac{1}{m}(\bm a' \cdot \bm \jmath + \frac{1}{2}\rho S|\bm v_a| \barbar{c}_0 \bm v_a \cdot \bm \jmath) \bm \jmath
\end{equation}
Recalling the relations from Eq.~\ref{eq:proof4} for the solutions of $|\bm T_r|$, $\gamma_{T,r}$, $\bm \imath_{2r}$ and $\bm k_{2r}$ and using the fact that $\bm \jmath_r$ is orthogonal to both $\bm a'$ and $\bm v_a$, Eq.~(\ref{eq:proof507}) becomes
\begin{align}\label{eq:proof508}
\begin{split}
	\dot{\bm v} = ~ & \bm a_r -\frac{1}{m}(\bm d \cdot (\bm \imath_2 - \bm \imath_{2r}) + |\bm T_r| cos(\gamma_{T,r}+\alpha_0) -|\bm T| \cos(\gamma_T + \alpha_0) ) \bm \imath_2 \\
	&-\frac{1}{m}(\bm e \cdot (\bm k_2- \bm k_{2r}) +|\bm T_r| \sin(\gamma_{T,r} + \alpha_0)-|\bm T| \sin(\gamma_T + \alpha_0) ) \bm k_2 \\
	&-\frac{1}{m}(\bm a' \cdot (\bm \jmath - \bm \jmath_r) + \frac{1}{2}\rho S|\bm v_a| \barbar{c}_0 \bm v_a \cdot (\bm \jmath - \bm \jmath_r)) \bm \jmath
\end{split}
\end{align}

Using the expressions of the pusher thrust $\bm T_{FW}$ and the collective thrust $\bm T_{MC}$ from Eq.~(\ref{eq:seperated-thrust}), as well as the relations between $(\bm \imath,\bm k)$ and $(\bm \imath_2, \bm k_2)$ from  Eq.~(\ref{eq:modified-body-axis}), one can verify that Eq.~(\ref{eq:proof508}) can be written as

\begin{align}\label{eq:proof509}
\begin{split}
	\dot{\bm v} = ~ & \bm a_r -\frac{1}{m}(\bm d \cdot (\bm \imath_2 - \bm \imath_{2r} )) \bm \imath_2- \frac{1}{m}(\bm e \cdot (\bm k_2- \bm k_{2r})) \bm k_2  \\
	&-\frac{1}{m}((\bm a'+ \frac{1}{2}\rho S|\bm v_a| \barbar{c}_0 \bm v_a ) \cdot (\bm \jmath - \bm \jmath_r))  \bm \jmath \\
	& +\frac{1}{m}  (|\bm T_{FW}|-|\bm T_{FW,r}|)\bm \imath   -\frac{1}{m} (|\bm T_{MC}|-|\bm T_{MC,r}|) \bm k
\end{split}
\end{align}

where $|\bm T_{FW,r}|$ and $|\bm T_{MC,r}|$ are the desired pusher and collective thrust.

Assuming that the propulsion systems track the thrust setpoints with relatively high bandwidths, this in turn implies that the tracking errors $(|\bm T_{FW}|-|\bm T_{FW,r}|)$ and $(|\bm T_{MC}|-|\bm T_{MC,r}|)$ converge to 0. Thus the last two terms in Eq.~\eqref{eq:proof509} converge to $\bm{0}$. 

 Also assuming that $|\bm v_a|$ is bounded due to energy dissipation in the air, and since the norm of the desired acceleration $|\bm a_r|$ is bounded by design (saturation of control terms), then $|\bm d|$, $|\bm e|$, and $|\bm a'+ \frac{1}{2}\rho S|\bm v_a| \barbar{c}_0 \bm v_a|$ are also bounded. Now if $\bm \imath_{2r}$, $\bm \jmath_{r}$ and $\bm k_{2r}$ are well defined, it suffices to design an attitude control law that makes $|\bm \imath_2 - \bm \imath_{2r}|$, $|\bm \jmath - \bm \jmath_{r} |$ and $|\bm k_2- \bm k_{2r}|$ converge to 0.

	We then assume that an attitude control law is designed to achieve the convergence of the body-fixed frame ${\mathcal B}=\{G, \bm \imath, \bm \jmath, \bm k  \}$  to a desired body-fixed frame ${\mathcal B_r}=\{G, \bm \imath_r, \bm \jmath_r, \bm k_r  \}$, with ${\mathcal B_r}$ being related to ${\mathcal B_{2,r}=\{G, \bm \imath_{2r}, \bm \jmath_{r}, \bm k_{2r}\}}$ according to the rotation equations~\eqref{eq:modified-body-axis}. Then, the convergence of $|\bm \imath_2 - \bm \imath_{2r}|$, $|\bm \jmath - \bm \jmath_{r} |$ and $|\bm k_2- \bm k_{2r}|$ to 0 follows immediately, which results in a translation dynamics of the form $\dot{\bm v}=\bm a_r + \bm o(t)$ and the achievement of the high-level position and speed objectives.

\subsection{Attitude control convergence analysis}\label{sec:app2}

Here, we give a proof for the convergence of the body-fixed frame $\mathcal{B}$ to the desired frame $\mathcal{B}_r$ using the angular velocity control of Eq.~(\ref{eq:att-cont2}). First let us derive the expression for the feed-forward term $\bm \omega_{ff}$. The angular velocities of the unitary vectors $\bm \jmath_r$ and $\bm k$ associated with the reference frame $\mathcal{B}_r$ are given by

\begin{equation}\label{eq:proof20}
\begin{cases}
\bm \omega_{\bm \jmath_r}&= \bm\jmath_r \times \frac{d \bm \jmath_r}{dt}\\
\bm \omega_{\bm k_r}&= \bm k_r \times \frac{d \bm k_r}{dt}\\
\end{cases}
\end{equation}

From these expressions, one can deduce that the angular velocity of the desired frame $\mathcal{B}_r$ can be computed as
\begin{equation}\label{eq:proof21}
\bm \omega_{ff} = \bm \omega_{\bm k_r} + (\bm \omega_{\bm \jmath_r} \cdot \bm k_r) \bm k_r
\end{equation}
Let $(\bm u_\theta, \tilde{\theta})$ denote the axis-angle representation of the rotation between the frames $\mathcal{B}$ and $\mathcal{B}_r$. Consider the following candidate Lyapunov function:
\begin{equation}\label{eq:proof22}
  V = \frac{1}{2} \tan^2(\frac{\tilde{\theta}}{2}) \geq 0
\end{equation}
Taking the derivative of $V$ gives
\begin{equation}\label{eq:proof23}
  \dot{V} = \frac{1}{2} \frac{\tan(\frac{\tilde{\theta}}{2})}{\cos^2(\frac{\tilde{\theta}}{2})} \dot{\tilde{\theta}} 
\end{equation}
By definition of the error angle $\tilde{\theta}$, one can write $\dot{\tilde{\theta}}=(\bm \omega_{ff} - \bm \omega) \cdot \bm u_\theta$. Replacing $\bm \omega$ by the control law of Eq.~(\ref{eq:att-cont2}) yields:
\begin{equation}\label{eq:proof24}
\dot{\tilde{\theta}} = - (k_{\bm \imath}( \bm{\omega_0}\cdot \bm \imath) \bm \imath + k_{\bm \jmath}( \bm{\omega_0}\cdot \bm \jmath) \bm \jmath + k_{\bm k}(\bm{\omega_0}\cdot \bm k) \bm k) \cdot \bm u_\theta
\end{equation}
From Rodrigues formula for the rotated vectors, one can deduce that
\begin{equation}\label{eq:proof25}
	 \bm{\omega_0} = (\bm \imath \times \bm \imath_r) + (\bm \jmath \times \bm \jmath_r) + (\bm k  \times \bm k_r)= 2 \sin(\tilde{\theta}) \bm u_\theta
\end{equation}
Replacing $\bm{\omega_0}$ in the expression of $\dot{\tilde{\theta}}$ gives:
\begin{equation}\label{eq:proof26}
\dot{\tilde{\theta}} = - 2 \sin(\tilde{\theta}) (k_{\bm \imath} (\bm u_\theta \cdot \bm \imath)^2 + k_{\bm \jmath} (\bm u_\theta \cdot \bm \jmath)^2 + k_{\bm k} (\bm u_\theta \cdot \bm k)^2)
\end{equation}
Let $\lambda=min(k_{\bm \imath},k_{\bm \jmath},k_{\bm k}) > 0$ denote the minimal control gain. Since $\bm u_\theta$ is a unitary vector, one can deduce that $\dot{\tilde{\theta}} \leq - 2 \lambda \sin(\tilde{\theta})$, which when replaced in the expression of $  \dot{V}$ gives:
\begin{equation}\label{eq:proof27}
  \dot{V} \leq -4 \lambda V
\end{equation}
Therefore, $V$ converges exponentially to 0, and almost global exponential stability of $\tilde{\theta}=0$ follows immediately. Note that orientations such that $\tilde{\theta}=\pi$ are unstable equilibria.

 \subsection{Local stability and perfomance analysis}\label{sec:app3}

\begin{figure}[H]
\centering
\includegraphics[width=.7\textwidth]{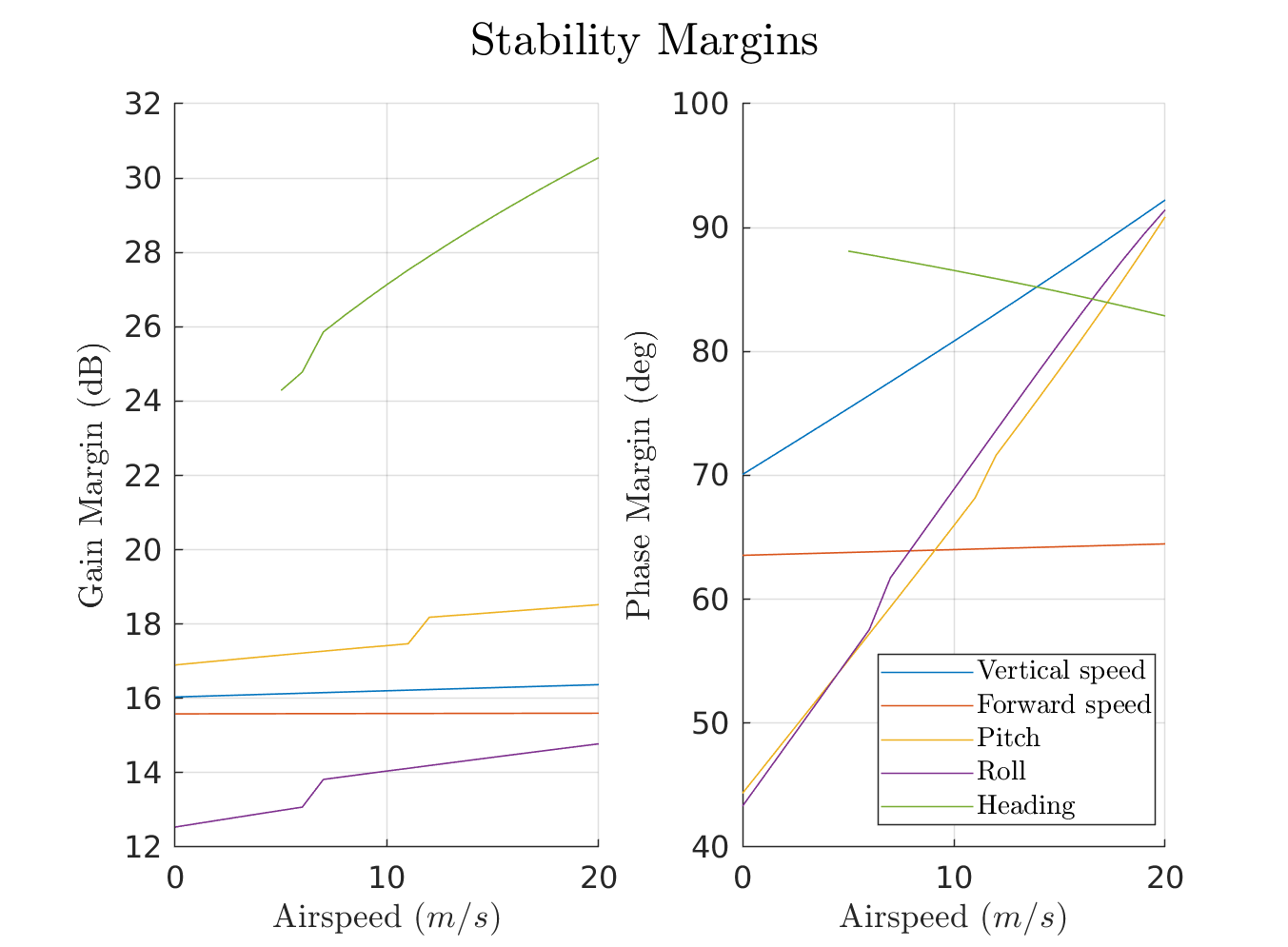}
\caption{Stability analysis.}
\label{fig:stab-analysis}
\end{figure}

\begin{figure}[H]
\centering
\includegraphics[width=.7\textwidth]{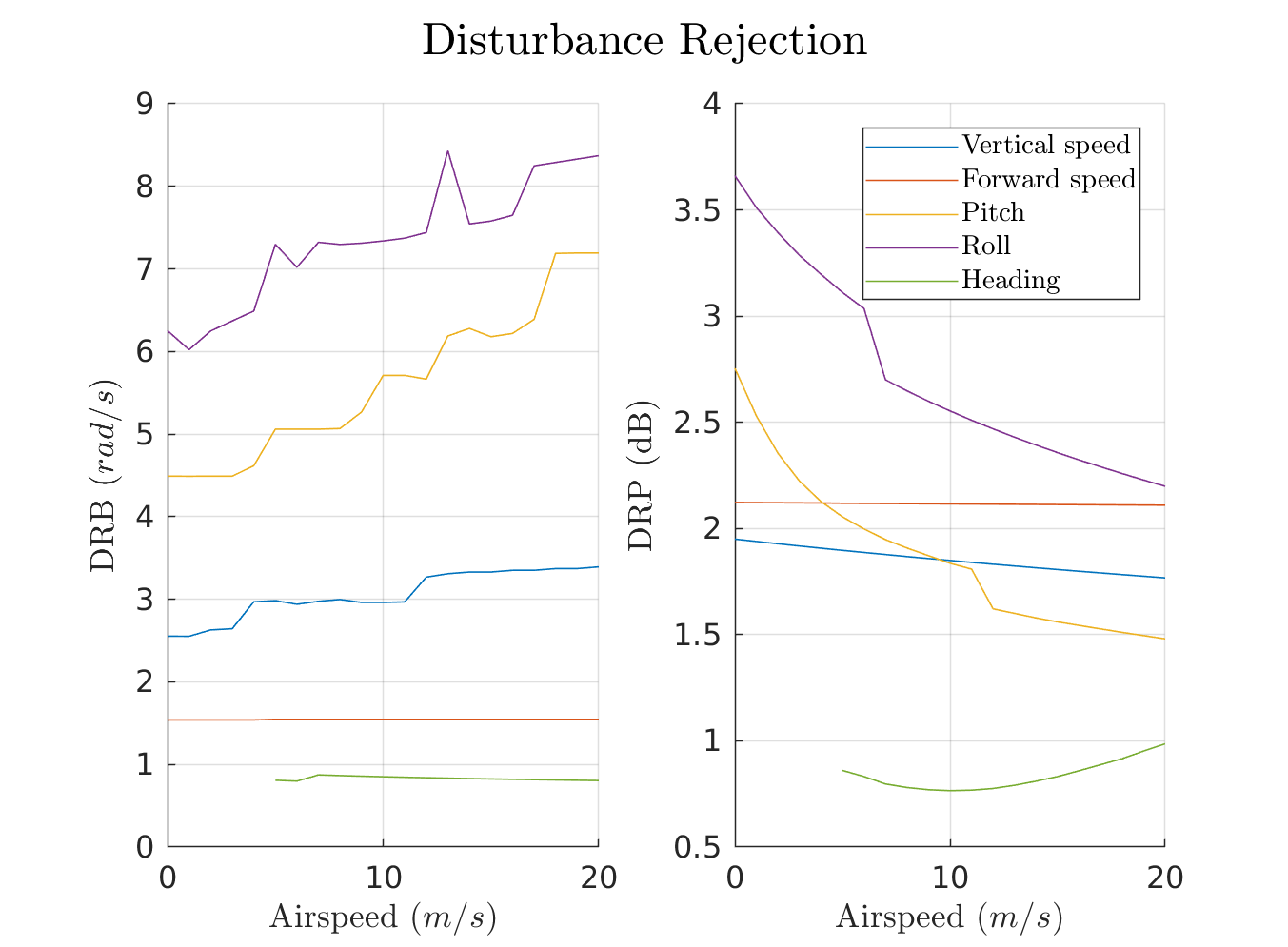}
\caption{Disturbance rejection analysis.}
\label{fig:dist-rej-analysis}
\end{figure}

Here, we present some stability, performance and disturbance rejection characteristics attained with the control parameters listed in Table \ref{tab:params}. The following analysis corresponds to locally linearized systems around equilibrium states along the transition maneuver. In particular, the nonlinear control terms are reduced to their locally linear equivalents. We also apply the reduced torque control expression in Eq.~\eqref{eq:ang-rate-cont2} rather than the full expression in Eq.~\eqref{eq:ang-rate-cont}. The different characteristics are shown in Figs.~\mref{fig:stab-analysis,fig:dist-rej-analysis} along the transition maneuver, as function of the airspeed ranging from $0$ to $20 m/s$. They correspond to decoupled dynamics for vertical speed, forward speed, pitch, heading and roll loops. Also note that because the targeted UAV lacks a side-slip sensor it was necessary to ensure that lateral Dutch roll modes are inherently stable.

\begin{figure}[H]
\centering
\includegraphics[width=.7\textwidth]{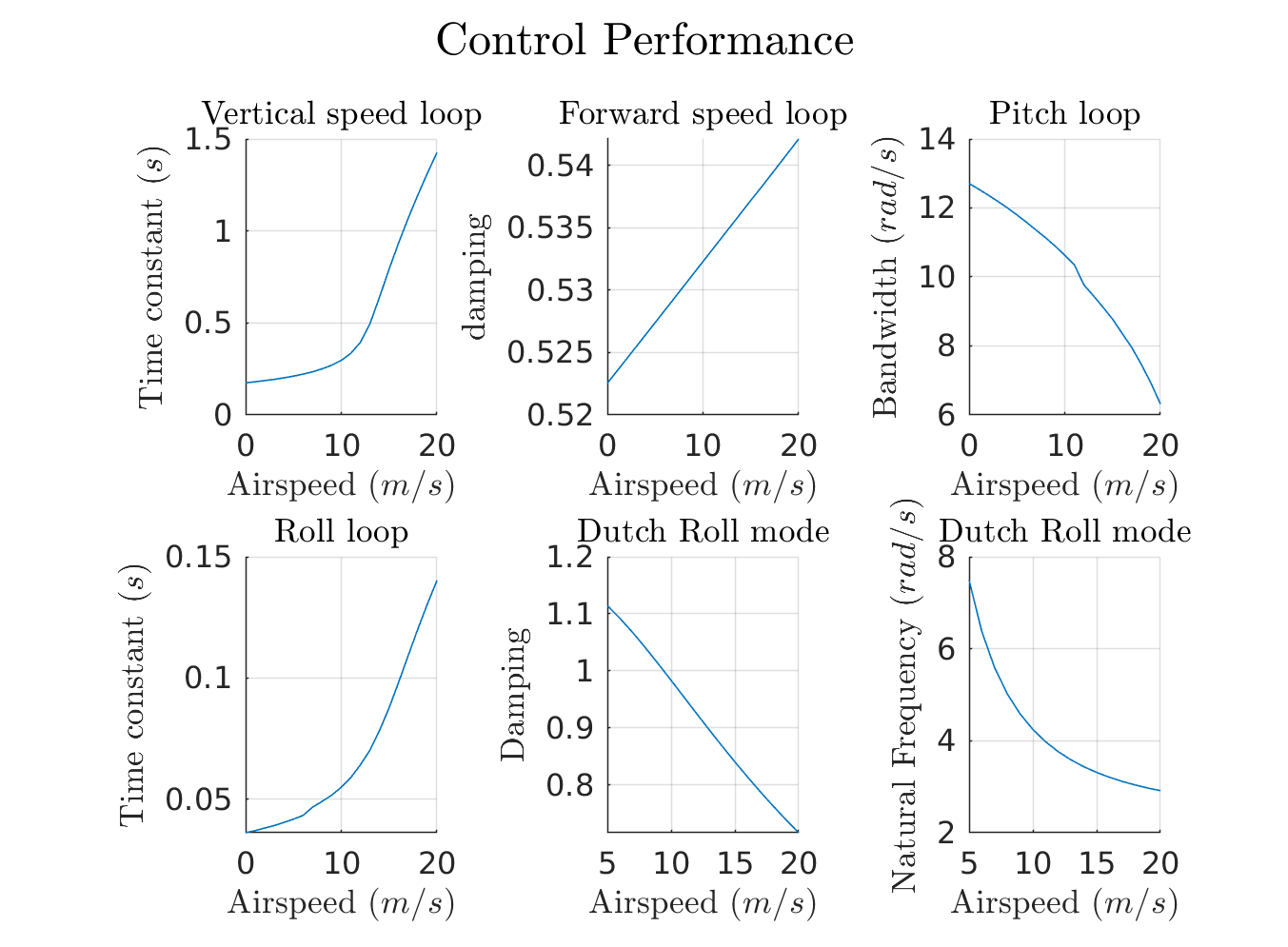}
\caption{Control performance analysis.}
\label{fig:cont-analysis}
\end{figure}

\section*{Acknowledgments}
This work was funded by Safran SA.

\bibliography{sample}

\end{document}